\documentclass[gmd,manuscript]{copernicus}

\usepackage{paralist}
\usepackage{xcolor}
\usepackage{overpic}
\usepackage{listings}
\newcommand{\refsec}[1]{Sect.~\ref{sec:#1}}

\usepackage{longtable}
\usepackage{hyperref}
\begin{document}

\title{Modular System for Shelves and Coasts (MOSSCO~v1.0) 
-- a flexible and multi-component framework for coupled coastal ocean ecosystem modelling}
\runningtitle{Modular System for Shelves and Coasts (MOSSCO)}

\Author[1]{Carsten}{Lemmen} 
\Author[1,5]{Richard}{Hofmeister}
\Author[2,3]{Knut}{Klingbeil}
\Author[4,5]{M.~Hassan}{Nasermoaddeli}
\Author[1]{Onur}{Kerimoglu}
\Author[2]{Hans}{Burchard}
\Author[4]{Frank}{K\"osters}
\Author[1]{Kai~W.}{Wirtz}

\affil[1]{Institute of Coastal Research, Helmholtz-Zentrum Geesthacht Zentrum 
f\"ur Material- und K\"ustenforschung,  21502~Geesthacht, Germany}
\affil[2]{Department of Physical Oceanography and Instrumentation, 
Leibniz-Institute for Baltic Sea Research, 18119~Rostock-Warnem\"unde, Germany}
\affil[3]{Department of Mathematics, Universit\"at Hamburg, 20146~Hamburg, Germany}
\affil[4]{Section Estuary Systems I, Bundesanstalt f\"ur Wasserbau, 22559~Hamburg, Germany}
\affil[4]{Landesbetrieb Stra{\ss}en, Br\"ucken und Gew\"asser, Freie und Hansestadt Hamburg, 20097 Hamburg, Germany}  
\affil[5]{Institute for Hydrobiology and Fisheries Science, Universit\"at Hamburg, 22767 Hamburg, Germany}  

\runningauthor{C. Lemmen et al.}
\correspondence{C. Lemmen\\ (carsten.lemmen@hzg.de)}

\received{}
\pubdiscuss{}
\revised{}
\accepted{}
\published{}

\firstpage{1}

\maketitle

\begin{abstract}
Shelf and coastal sea processes extend from the atmosphere through the water
column and into the sea bed. These processes are driven by physical, chemical,
and biological interactions at local scales, and they are influenced by transport
and cross strong spatial gradients.  The linkages between domains and many
different processes are not adequately described in current model systems. Their
limited integration level in part reflects lacking modularity and flexibility; this
shortcoming hinders the exchange of data and model components and has
historically imposed supremacy of specific physical driver models. 
We here present the Modular System for Shelves and Coasts (MOSSCO, 
\href{http://www.mossco.de}{http://www.mossco.de}), a novel domain and process
coupling system tailored -- but not limited -- to the coupling challenges of and
applications in the coastal ocean.  MOSSCO builds on the existing coupling
technology Earth System Modeling Framework and on the Framework
for Aquatic Biogeochemical Models, thereby creating a unique level
of modularity in both domain and process coupling; the new framework adds  rich
metadata, flexible scheduling, configurations that allow several tens of models
to be coupled, and tested setups for coastal coupled applications.
That way, MOSSCO addresses the technology needs of a growing marine coastal Earth System
community that encompasses very different disciplines, numerical tools, and
research questions. 
\end{abstract}

\introduction
Environmental science and management consider ecosystems as their primary
subject, i.e. those systems where the organismic world is fundamentally linked to 
the physical system surrounding it; there exist neither unequivocally defined  
spatial nor processual boundaries between the components of an ecosystem
\citep{Tansley1935}.  Consequently, holistic approaches to  ecological research
\citep{Margalef1963}, to biogeochemistry \citep[][originally 1926]{Vernadsky1998}
and to environmental science in general \citep{Lovelock1974} have been called for. 

The need for systems approaches  is perhaps most apparent in coastal research.
Shelf and coastal seas are described by components from different spatial 
domains: atmosphere, ocean, soil; and they are driven by a manifold of interlinked
processes: biological, ecological, physical, geomorphological,  amongst others. 
Crossing these domain and process boundaries, the dynamics of suspended
sediment particles (SPM, see Table~\ref{tab:abbreviations} for abbreviations)
and of living particles, or the interaction between water attenuation and phytoplankton
growth, for example, are both scientifically challenging and relevant for the ecological
state of the coastal system \citep[e.g.,][]{Shang2014,Maerz2011,Azhikodan2016}.

For historic and practical reasons, the representation of the coastal ecosystem in
numerical models has been far from holistic. Most often, ecological and
biogeochemical processes are described in modules that are tightly coupled to
one hydrodynamic model.  For example, the Pelagic Interactions Scheme for
Carbon and Ecosystem Studies \citep[PISCES,][]{Aumont2015} has been
integrated into the Nucleus for European Modeling of the Ocean
\citep[NEMO,][]{VanPham2014}. Or, the Biogeochemical Flux Model (BFM) has
been integrated in the Massachusetts Institute of Technology Global 
Circulation Model (MITgcm) \citep{Cossarini2017}.  These tight couplings not only exclude
important processes at the edges of or beyond the pelagic domain, they also lack flexibility
to exchange or to test different process descriptions.    

To stimulate the development, application and interaction of ecological and
biogeochemical models independently of a single host hydrodynamic model,
\citet{Bruggeman2014} presented the Framework for Aquatic Biogeochemical
Models (FABM), which serves as an intermediate layer between the
biogeochemical zero-dimensional process models and the three-dimensional
geophysical environment models.  FABM has been implemented in the Modular
Ocean Model \citep[MOM,][]{Bruggeman2014}, NEMO, the Finite Volume
Coastal Ocean Model \citep[FVCOM,][]{Cazenave2016}, or the General
Estuarine Transport Model \citep[GETM,][]{Kerimoglu2017}. With more than
20~biogeochemical and ecological models included, FABM has enabled marine
ecosystem researchers to describe the system's many aquatic processes.

The process-oriented modularity realized within FABM, however, lacks the means to
describe cross-domain linkages. Historically rooted in atmosphere--ocean circulation models 
\citep{Manabe1969}, the coupling of earth domains is the standard concept in 
Earth System Models (ESM). Domain coupling is also a major challenge in coastal modelling 
and has been used, for example, in the Coupled Ocean-Atmosphere-Wave-Sediment Transport
\citep[COAWST,][]{Warner2010} system.  COAWST comprises  the Regional
Ocean Modeling System (ROMS) with a tightly  coupled sediment transport model, the
Advanced Research  Weather Research and Forecasting (WRF)
atmospheric model, and the Simulating Waves Nearshore (SWAN) wave model.
Each of the components in domain coupling is usually a self-sufficient model that is
run in a special ``coupled mode''.  Interfacing to other components is done via
generic coupling infrastructure, such as the Flexible Modeling System
\citep[FMS,][]{Dunne2012}, the Model Coupling Toolkit 
\citep[MCT,][]{Warner2008ems} and/or  the Ocean Atmosphere Sea Ice Soil (OASIS)
coupler \citep{Valcke2013}, or the Earth System Modeling Framework (ESMF, 
\citealt{Hill2004}, see, e.g., \citealt{Jagers2010} for an intercomparison of coupling
technologies).  Typically, three to five domain
components are coupled through one of these technologies \citep{Alexander2015}.  

The differentiation between domain and process coupling is not a technical necessity:
A typical domain coupling software like ESMF can indeed be used to
couple processes:  with the Modeling, Analysis and Prediction Layer
\citep[MAPL,][]{Suarez2007}, the  Goddard Earth Observing System version~5
(GEOS-5) encompasses 39~process models coupled  hierarchically through
ESMF; development of these modules, however, is strictly regulated within the developing laboratory.
Vice versa, a typical process coupling infrastructure like the  Modular Earth Submodel System
\citep[MESSy,][]{Joeckel2005} has been proposed to link processes across domains,
but so far includes mostly atmospheric processes. 

Currently, there is no coastal modelling environment that enables a  modular and flexible
process (model) integration and cross-domains coupling at the same time, and that is
open to a larger community of independent biogeochemical and ecological scientists.
The underlying long-term goal for increasingly holistic model systems conflicts with the
evolving and diverse research needs of individual scientists or research groups to address
very specific problems; it remains difficult to link up-to-date research that is delivered at
the (local) process scale to the Earth System scale.
Thus we here present the  Modular System for Shelves and Coasts (MOSSCO,
\href{http://www.mossco.de}{www.mossco.de}), a novel domain and process coupling
system tailored -- but not limited -- to the coupling challenges of and applications in the
coastal ocean.  This new system builds on the
flexibility of FABM and on the infrastructure provided by ESMF with its cross-domain
and many-component hierarchical capability.  We here present the major design ideas of MOSSCO
and briefly demonstrate its usability in a series of coastal applications. 

\section{MOSSCO concepts}

The modularity and coupling concepts proposed in this paper
elaborate the design of a novel software system that emphasizes the needs of
researchers who want to  make maximum use of their existing knowledge in a  specific
field (e.g., geomorphology or marine ecology) but wish to conduct integrative research
in a wider and flexible context. In strengthening modularity \emph{sensu} independence
of specific physical drivers, the new concept should, in addition to addressing the
problems listed above, support
\begin{inparaenum}[(1)]
\item liaisons  between traditionally separated modelling communities (e.g., coastal
engineers, physical oceanographers and biologists),
\item inter-comparison  studies of, e.g., physical, geological, and biological modules, and
\item up-scaling studies where models developed at the  laboratory scale in a
non-dimensional context are applied to regional, global and Earth System scales.
\end{inparaenum}

The design of MOSSCO is application-oriented and driven by the demands for enabling and improving integrated
regional coastal  modelling. It is targeted towards building coupled systems
that support decision making for local policies implementing the European Union Water
Framework Directive (WFD) and Marine Strategic Planning Directive (MSPD).  From a design 
point of view we envisioned a system that is foremost flexible and equitable.

\begin{description}
\item[Flexibility] means that the system itself is able to deal on the one hand 
with a diverse small or large  constellation of coupled model components and on
the other hand with different orders of magnitude of spatial and temporal
resolutions; it is able to deal equally well with zero-, one-, two- and three-dimensional
representations of the coastal system. Flexibility implies the capability to encapsulate
also existing legacy models to create one or more different ``ecosystems'' of models.
This feature should allow seamless replacement of individual model components, which
is an important procedure in the continual development of integrated systems. Flexibly
replacing components finally creates a test-bed for model intercomparison studies.

\item[Equitability] means that all models in the coupled framework are treated as 
equally important, and that none is more important than any other.  This principle
dissolves the primacy of the hydrodynamic or atmospheric models as the
central hub in a coupled system.   Also, data components are as important as process 
components or model output; any de facto difference in model importance should be
grounded on the research question, and not on technological legacy.  As complexity
grows by coupling
more and more models, this equitability also demands that experts in one particular
model can rely on the functionality of other components in the system without having
to be an expert in those models, as well. 
\end{description}

The equitability design extends to participation:  contributions to the development
of  components or the coupling framework itself is allowed and
encouraged.  Anyone can use and modify the coupled framework or parts of both
in a legal sense by open source licensing, and in an accessibility sense through
template codes and extensive documentation.

\subsection{Wrapping legacy models -- first steps in  \texttt{PARSE}}

As MOSSCO is built around the ESMF hierarchy of components, any existing code
that can be wrapped in an ESMF component can be a component in MOSSCO, too.  
The ESMF user guide \citep{ESMF2013usr} suggests a best 
practice method \texttt{PARSE} to achieve this componentization of a legacy code.  

\begin{description}
\item[\texttt{P}]repare the user code by splitting it into three phases that initialize, run and finalize a model;
\item[\texttt{A}]dapt the model's data structures by wrapping them in ESMF infrastructure like states and fields;
\item[\texttt{R}]egister the user's initialize, run, and finalize routines through ESMF;
\item[\texttt{S}]chedule data exchange between components;
\item[\texttt{E}]xecute a user application by calling it from an ESMF driver.
\end{description}

This \texttt{PARSE} concept allows a smooth transition from a legacy model to an ESMF component. In this concept, the first 
three steps have to be performed on the model side, and the latter two on the framework side and have been taken care 
of by the MOSSCO coupling layer. The \emph{preparation} of the code is independent of the use of ESMF and provides 
the basic couplability of the model; many existing models already implement this separation into initialize, run, and finalize phases, 
either structurally or more formally by implementing a Basic Model Interface \citep[BMI,][]{Peckham2013}.  For the run 
phase, it is mandatory that this phase refers to a single model timestep and not to the entire run loop.

The \emph{adaption} of a model's internal structures to ESMF consists of technically wrapping data into ESMF 
communication objects, and in providing sufficient metadata for communication.  Among these are grid definition and 
decomposition, units and semantics of data, optimally following a metadata scheme like the widespread Climate and 
Forecast \citep[CF,][]{Eaton2011} or the more bottom-up Community Surface Dynamics Modeling System 
(CSDMS, following a scheme like  
object + [ operation ] + quantity, \citealt{Peckham2014iemss}).  Both are currently being included in the emerging 
Geoscience Standard Names Ontology (GSN, \href{http://geoscienceontology.org}{geoscienceontology.org}).  

ESMF provides the interfaces for models written in either the Fortran or C~programming languages; data 
arrays are bundled together with related metadata in ESMF field objects.  All field objects from components are then 
bundled into exported and imported ESMF state objects to be passed between components.  As a third step, the
ESMF \emph{registration} facility needs to be added to a user model; this step is achieved by using template code from 
any one of the examples or tutorials provided with ESMF.  The second and third step (\emph{adapt} and 
\emph{register}) are  typical tasks of what  \citet[][]{Peckham2013} refers to as a component model interface (CMI);
it  is very similar between models (and thus easily accessible from template code) and targets the interface of a 
specific coupling framework.  

\begin{figure}
\includegraphics[width=.5\hsize]{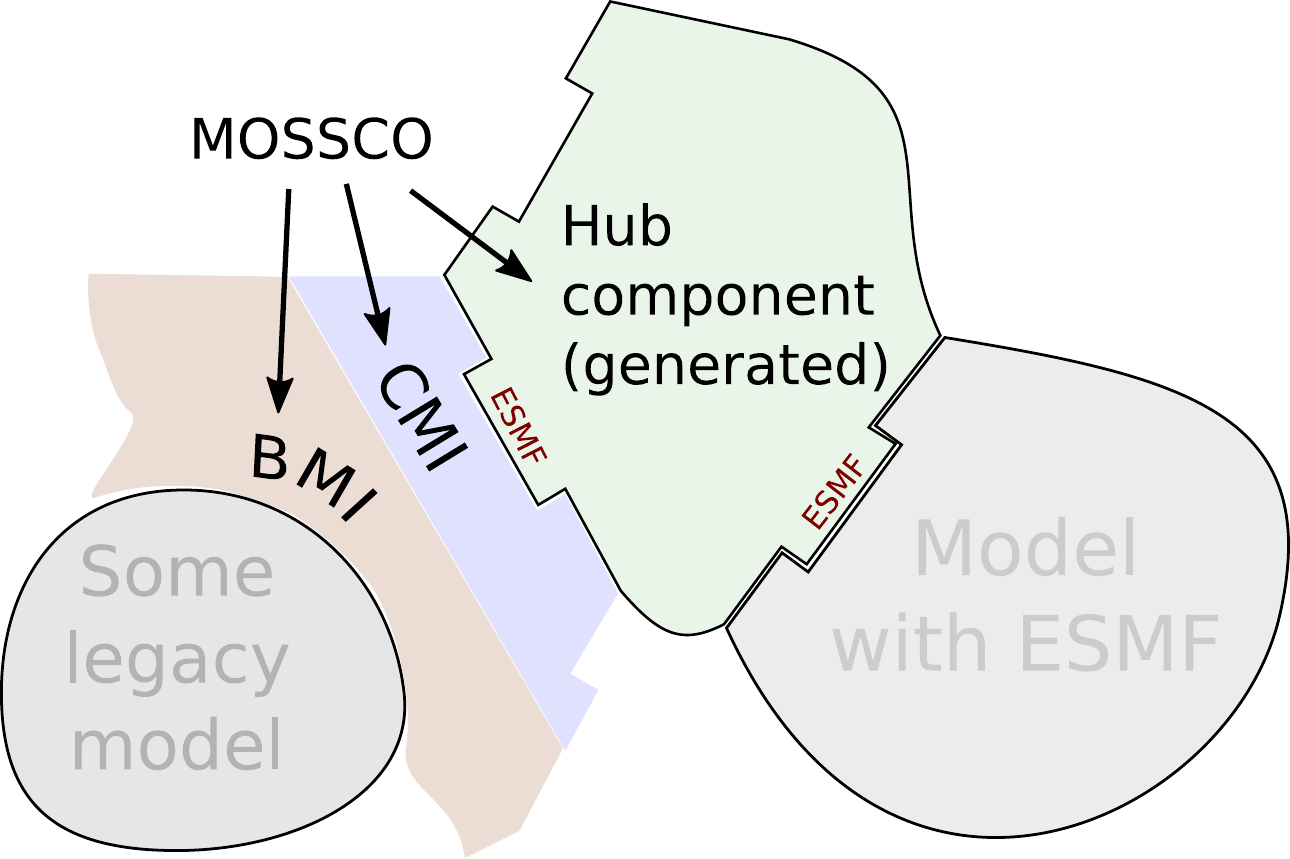}
\caption{MOSSCO's adoption of legacy code follows the two-layer paradigm of BMI/CMI (basic model interface/
component model interface) suggested by \citet{Peckham2013}. An existing 
legacy code (illustrated by ``some model'') is enhanced by model-specific code that exhibits basic coupling functionality (``BMI'') and is 
framework agnostic.  In a second step, a component (``CMI'') is added, that uses the BMI interface in the specific 
application programming interface of the coupling framework.  In addition to model interfaces that can be used in MOSSCO-independent
contexts, MOSSCO provides coupling concepts and working implementations for 
coupled applications.}
\label{fig:bmicmi}
\end{figure}

MOSSCO contains CMIs for ESMF in all of its provided components (Fig.~\ref{fig:bmicmi}).  The 
current naming scheme  follows the CF convention for standard names except for quantities that are not defined by CF; 
these names (often from biological processes) are modeled onto existing CF standard names as much as possible.  
MOSSCO also allows the specification within other naming schemes and includes a name matching 
algorithm to mediate between different schemes. For future development, adoption of the GSN ontology is 
foreseen.  

\subsection{Scheduling in a coupled system -- the ``S'' in  \texttt{PARSE}}

MOSSCO adds onto ESMF a scheduling system (corresponding to the fourth step in
\texttt{PARSE}) that  calls  the different phases of participating coupled models.  The
coupling time step duration of this new scheduler relies on the ESMF concept of alarms
and a user specification of pairwise coupling intervals between models. The scheduler
minimizes calls to participating models by flexibly adjusting time step duration to the
greatest common denominator of coupling intervals pertinent to each coupled model.
Upon reading the user's coupling specification, 
\begin{inparaenum}[(i)]
\item models are initialized in random order but with consideration of special initialization dependencies set by the user;
\item a list of alarm clocks is generated that considers all pairwise couplings a model is involved in;
\item special couplers associated with a pairwise coupling are executed;
\item the scheduler then tells each model to run until that model reaches its next alarm time;
\item  advancing of the scheduler to the minimum next alarm time repeats until the end of the simulation.
\end{inparaenum}

The MOSSCO scheduler allows for both sequential and for concurrent coupling of model components, or a hybrid 
coupling mode.  In the concurrent mode, components run at the same time on different compute elements; in the 
sequential mode, components are executed one after another.  Recently,  \citet{Balaji2016} demonstrated how a hybrid 
coupling mode and fine granularity could be used to increase the performance of a system that consists of both highly 
scalable and less scalable components: In their system, they ran an ocean component concurrently with the radiation 
code of the atmosphere sequentially to all other atmospheric process components.  

\begin{figure*}
\centering
\begin{overpic}[grid=false,viewport=0 135 280 370,clip=,width=.5\hsize]{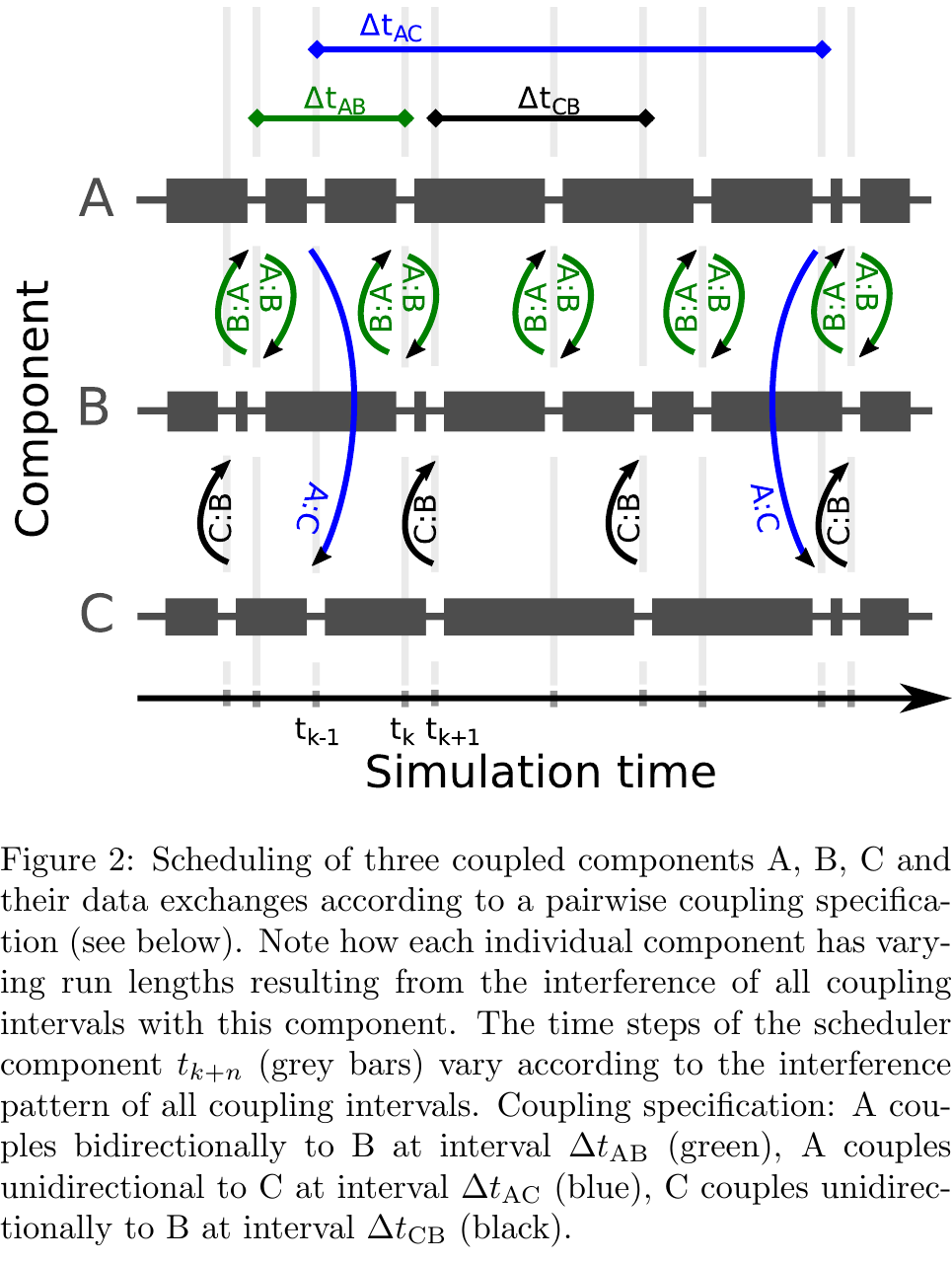}
\end{overpic}
\caption{Scheduling of three coupled components A, B, C and their data exchanges according to a pairwise coupling specification (see below). Note how each individual component has varying run lengths resulting from the interference of all coupling intervals with this component.  The time steps of the scheduler component  $t_{k+n}$ (grey bars) vary according to the interference pattern of all coupling intervals.  Coupling specification: A couples bidirectionally to B at interval $\Delta t_\mathrm{AB}$ (green), A couples unidirectional to C at interval $\Delta t_\mathrm{AC}$ (blue), C couples unidirectionally to B at interval $\Delta t_\mathrm{CB}$ (black).}
\label{fig:scheduling}
\end{figure*}

For both concurrent and sequential modes, the MOSSCO scheduler runs the connectors and mediator components
that exchange the data before the components are run, i.e. computations are performed on data with the same timestamp.  For sequential mode, the 
coupling configuration allows also a scheme where consecutive components rely always on the most recently 
calculated data from all other components (Fig.~\ref{fig:scheduling}, see also Sect.~\ref{sec:linkcopy}).

\subsection{Deployment of the coupled system -- the ``E'' in  \texttt{PARSE}}

MOSSCO provides a Python-based generator that dynamically creates an ESMF driver component
in a star topology that then acts as the scheduler for the coupled system.  This generator reads a text-based
specification of pairwise couplings (including the coupling interval, dependencies and instantiation under
different names) and generates a Fortran source file that represents the scheduler component.  The
generator takes care of compilation dependencies of the coupled models, and of coupling dependencies, such
as grid inheritance; in addition to the basic init--run--finalize BMI scheme, it also honors multi-phase
initialization (as in the National Unified Operational  Prediction Capability, NUOPC, ESMF extension) and a restart phase.

A MOSSCO command line utility provides a user-friendly interface to generating the scheduler,
(re-)compiling all source codes into an executable and submitting the executable to a multi-processor system,
including different high-performance computing (HPC) queueing implementations; this is the fifth step in 
\texttt{PARSE}.  MOSSCO has been successfully deployed at several
national HPC centers, such as the Norddeutsche Verbund f\"ur Hoch- und H\"ochstleistungsrechnen (HLRN), 
the  German Climate Computing Center (DKRZ), or the J\"ulich Supercomputing Centre  (JSC); equally,
MOSSCO is currently functioning on a multitude of Linux and macOS laptops, desktops and multiprocessor
workstations using the same MOSSCO (bash-based) command line utility on all platforms.  

The MOSSCO coupling layer is coded in Fortran while most of the supporting structure is coded in
Python and partially in Bash shell syntax.  The system requirements are a Fortran 2003 compliant
compiler, the CMake build system, the Git distributed version control system, Python with YAML
support (version~2.6 or greater), a Network Common Data Form \citep[NetCDF,][]{Rew1990}  library, and
ESMF (version~7 or greater).   For parallel applications, a Message Passing  library (e.g., OpenMPI) 
is  required. Many HPC centers have toolchains available that already meet all of
these requirements.  For an individual user installation, all requirements can be taken care of with
one of the package managers distributed with the operating system, except for the installation of
ESMF, which needs to be manually installed; MOSSCO provides a semiautomated tool for helping
in this installation of ESMF. 

\section{MOSSCO components and utilities}

\begin{figure*}
\centering
\begin{overpic}[grid=false,viewport=10 80 258 312,clip=,width=.5\hsize]{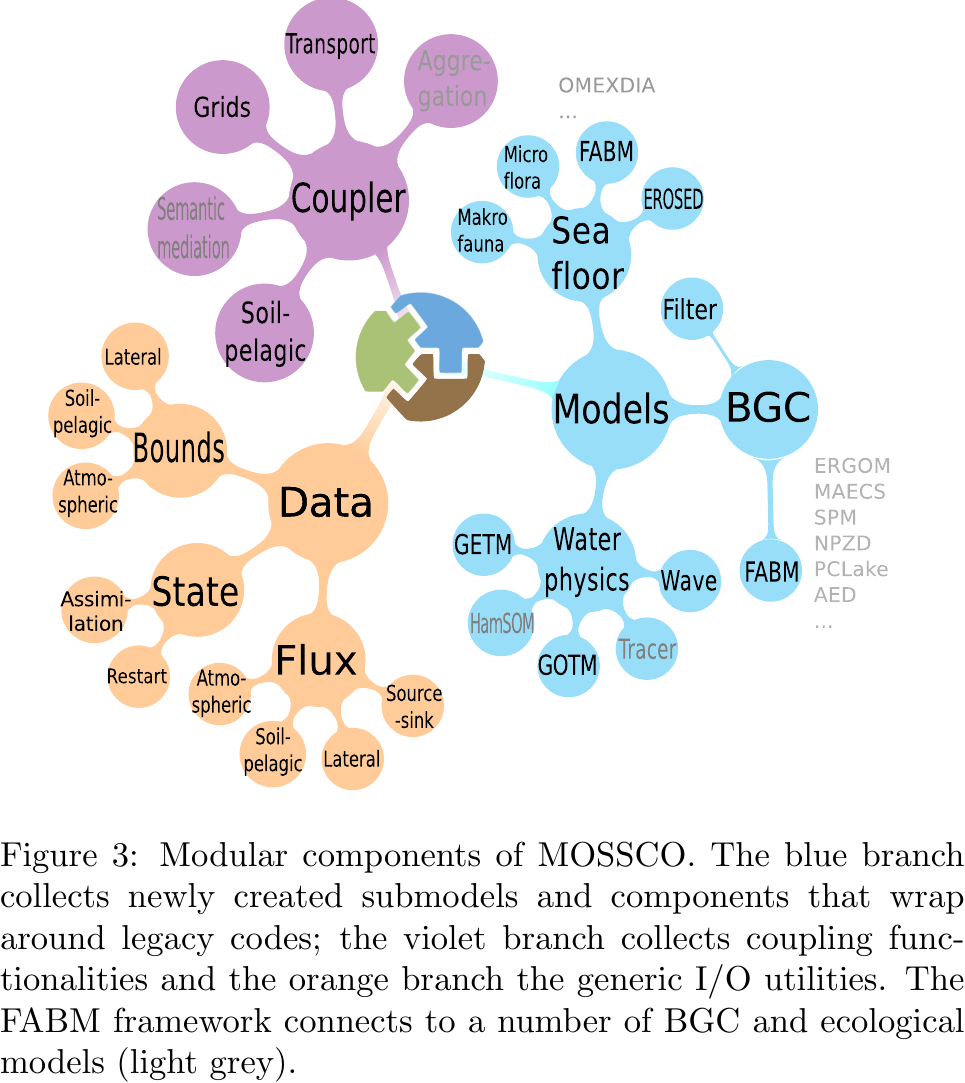}
\end{overpic}
\caption{Modular components of MOSSCO.  The blue branch collects newly created submodels
and components that wrap around legacy codes; the violet branch collects coupling functionalities
and the orange branch the generic I/O utilities.}
\label{fig:functionalities}
\end{figure*}

Driven by user needs, MOSSCO currently entails utilities for I/O, an extensive model library, and
coupling functionalities (Fig.~\ref{fig:functionalities} and Table~\ref{tab:components}).   As a utility
layer on top of ESMF, MOSSCO also extends the  Application Programming Interface (API) of
ESMF by providing convenience methods to facilitate the handling of time, metadata (attributes),
configuration, and to unify the provisioning and transfer of scientific data across the coupling
framework.  The use of this utility layer is not mandatory;  any ESMF based component can be
coupled to the MOSSCO provided components without using this utility layer.

One of the major design principles of MOSSCO is  seamless scaling from zero-dimensional to three-dimensional spatial
representations, while maintaining the coupling configuration to the maximum extent possible.
This design principle builds on the dimensional-independency concept of FABM achieved by local description of processes
(often referred to as a box model),  where the dimensionality is defined by the hydrodynamic model to which FABM is 
coupled; MOSSCO generalizes this concept to enable especially the developers of new biological and chemical models to scale up from a
box-model (zero-dimensional) to a water-column (one-dimensional), sediment plate or a vertically resolved transect
(two-dimensional), and a full atmosphere or ocean (three-dimensional) setup.
As a concrete example, upscaling of the novel Model for Adaptive Ecosystems in Coastal Seas \citep[MAECS,][]
{Wirtz2016,Kerimoglu2017} has been developed in this way going forth and back between the lab
scale and the regional coastal ocean scale. 

All utility functions and components, especially the generic I/O facilities from MOSSCO, are able to
handle data of any spatial dimension.  MOSSCO communicates different dimensional information
by grid inheritance: components that implement this are able to obtain the spatial information from another model in the coupled context.  
Usually (but not necessarily) biological and chemical models inherit the spatial configuration from a hydrodynamic model; 
equally well, this information can be obtained from data in standardized grid description formats like Gridspec 
\citep{Balaji2007} or the Spherical Coordinate Remapping and Interpolation Package  \citep[SCRIP,][]{Jones1999mwr}.

\subsection{Model library: Basic Model Interfaces for scientific model components}

The model library (right branch in Fig.~\ref{fig:functionalities}) includes new models (e.g., for filter
feeders and  surface waves) and wrappers to legacy models and frameworks such as FABM or
GETM.  Some of these wrappers are  under development (e.g., Hamburg Shelf Ocean Model
\citep[HamSOM,][]{Harms1997} and a Lagrangian particle tracer model). Here, we briefly document
the model collection in particular with respect to their preparation and functioning within the new
coupling context. 

\subsubsection{Pelagic ecosystem component}\label{sec:fabm_pelagic}

The pelagic ecosystem component (\texttt{fabm\_pelagic\_component}) collects (mostly biological) 
process models for aquatic systems. This component makes use of the Framework for Aquatic
Biogeochemical Models  \citep[][]{Bruggeman2014}.  FABM is a coupling layer to a multitude
of biogeochemical models which provide the source-minus-sink terms for variables, their vertical
local movement (e.g., due to sinking or active mobility), and diagnostic data. Each model variable 
is equipped with meta data, which is transferred by the ecosystem component into
ESMF field names and attributes. Similarly, the forcing required by the biogeochemical models
is communicated within the framework and linked to FABM. The pelagic ecosystem component
includes a numerical integrator for the boundary fluxes and local state variable dynamics. 
Advective and diffusive transport are not part of this component but are left to the hydrodynamical
model through the \texttt{transport\_connector} (\refsec{transport}). The close connection between
transport  and the pelagic ecosystem requires grid inheritance by the pelagic ecosystem
model from the hydrodynamic model component.

Many well known biogeochemical process models have been coded in the FABM standard by
various institutes, such as the European Regional Seas Ecosystem Model
\citep[ERSEM,][]{Butenschoen2016}, ERGOM \citep{Hinners2015}, PCLake \citep{Hu2016} and
the Bottom RedOx Model \citep[BROM,][]{Yakushev2017}.  All biogeochemical models complying
with the FABM standard can equally be used in MOSSCO, while retaining their functionality.

\subsubsection{Sediment/soil component}\label{sec:fabm_sediment}

The sediment component \texttt{fabm\_sediment\_component}  hosts (mostly biogeochemical)
process descriptions for aquatic soils. To allow efficient coupling to a pelagic ecosystem, the sediment 
component inherits a horizontal grid or mesh from the coupled system and adds its own vertical coordinate, a number 
of layers of horizontally equal height for the upper soil (typical domain heights range from 10--50\,\unit{cm}). State 
variables within the sediment are defined through the FABM framework within the 3D grid or mesh in the sediment. 
As in the pelagic ecosystem component, the state variables, meta data, diagnostics and forcings are 
communicated via the ESMF framework to the coupled system. The sediment component is the numerical integrator for 
the tracer dynamics within each sediment column in the horizontal grid or mesh, including diffusive transport, 
driven by molecular diffusion of the nutrients or bioturbative mixing.
Additionally, a 3D variable porosity defines the fraction of pore water as part of the bulk 
sediment, while all state variables are measured per volume pore water in each cell.  FABM's infrastructure of state 
variable properties is used to label the new Boolean property \texttt{particulate} in FABM models to define whether a state 
variable belongs to the solid phase within the domain.
A typical model used in applications of the sediment component is the biogeochemical model of the 
Ocean Margin Exchange Experiment OMEXDIA \citep[][]{Soetaert1996}; a version of this model
with added phosphorous cycle is contained  in the FABM model library as OMEXDIA\_P \citep{Hofmeister2014iche}. 

\subsubsection{1D Hydrodynamics: General Ocean Turbulence Model (GOTM))}\label{sec:gotm}

The  General Ocean Turbulence Model \citep[GOTM,][]{Burchard1999,Burchard2006jms} is
a one-dimensional  water column model for hydrodynamic and thermodynamic processes
related to vertical mixing. MOSSCO provides a component for GOTM and a component
hierarchy that considers a coupled GOTM with internally coupled FABM within one
component (\texttt{gotm\_fabm\_component}), as many existing available model setups rely on the direct
coupling of FABM to GOTM.  This way, the modularization -- taking a coupled GOTM/FABM apart and
recoupling it through the MOSSCO infrastructure, can be verified; the encapsulation of GOTM is implemented
in the \texttt{gotm\_component}.

\subsubsection{3D Hydrodynamics: General Estuarine Transport Model (GETM)}\label{sec:getm}
MOSSCO provides an interface to the 3D coastal ocean model GETM \citep{Burchard2002ec}.
GETM solves the Navier-Stokes Equations under Boussinesq approximation, optionally
including the nonhydrostatic pressure contribution \citep{Klingbeil2013}. A direct interface to
GOTM (see Section~\ref{sec:gotm}) provides state-of-the-art turbulence closure in the vertical.
GETM supports horizontally curvilinear and vertically adaptive meshes
\citep{Hofmeister2010,Graewe2015}; the different mesh topologies need to be made accessible
as ESMF grid objects. Typically,  the GETM component exports its grid and subdomain decomposition to the
coupled system where the spatial and parallelization information is inherited by other 
components. The interface to GETM is provided by the 
\texttt{getm\_component};  any model coupled to GETM via the transport component can
have its state variables conservatively transported by GETM (see Section~\ref{sec:transport}).

\subsubsection{Model components for erosion, sedimentation, and their biological alteration}\label{sec:erosed}

The erosion/sedimentation routines of the Deltares Delft3D model \citep[EROSED,][]
{VanRijn2007} were encapsulated in a MOSSCO component.   EROSED uses a Partheniades--Krone equation 
\citep{Partheniades1965} for calculating the net sediment flux of cohesive sediment at the water--sediment interface for 
multiple SPM size classes.  MOSSCO's \texttt{erosed\_component} always uses the current version of 
EROSED maintained by Deltares, and with the help of subsidiary infrastructure isolates the EROSED code from the deeply intertwined 
dependencies of the original implementation.  The functionality of this erosion and sedimentation component is described in more 
detail by \citet{Nasermoaddeli2014ich}.

Flow and sediment transport can be affected by the presence of benthic organisms in many ways. Protrusion of 
benthic animals and macrophytes in the boundary layer changes the bed roughness and thus the bed shear stress and 
consequently the sediment transport. The erodibility of sediment can be modified by the mucus produced by benthic 
organisms; the 
erodibility of the upper bed sediment can be altered by bioturbation generated by macrofauna \citep{DeDeckere2001}.  
In the \texttt{benthos\_component}, these biological effects of microphytobenthos and of benthic macrofauna on 
sediment erodibility and critical bed shear stress are parameterized. The benthos effect model is described in detail by 
\citeauthor{Nasermoaddeli2017} (submitted to Estuarine, Coastal, and Shelf Science).

\subsubsection{Filter feeding model}\label{sec:filtration}

The generic filtration component describes the instantaneous filtration by filter feeders within the water column. The biological 
filtration model follows \citet{Bayne1993} and describes the filtration rate as a function of food supply; it  can be adapted 
to different species of filter feeders and was recently applied to describing the ecosystem effect of blue mussels on 
offshore wind farms as the \texttt{filtration\_component} of MOSSCO (\citeauthor{Slavik2017}, to be submitted to Hydrobiologia).
The filtration model uses an arbitrary chemical species or compound, say phytoplankton carbon as the ``currency'' for 
processing.  The amount of ambient phytoplankton carbon concentration is sensed by the model organisms and it is 
filtered along with the other nutrients (in stoichiometric proportion) out of the environment, creating a sink term for 
subsequent numerical integration in the pelagic ecological model.  

\subsubsection{Wind waves}\label{sec:waves}

A simple wind wave model is part of the MOSSCO suite. Based on the parameterization by \citealt{Breugem2007},
significant wave height and peak wave period are estimated in terms of local water depth, wind speed and fetch length.
This wave data enable the inclusion
of wave effects especially for idealized 1D water column studies, e.g. the consideration of erosion processes
due to wave-induced bottom stresses.
Coupling to 3D ocean models and the calculation of additional wave-induced momentum forces there, following either the Radiation stress or
Vortex Force formulation \citep{Moghimi2013a}, is possible as well.
For the inclusion of wave--wave or wave--current interaction in realistic 3D applications,
the coupling to a more advanced third generation wind wave model like SWAN, WaveWatch~III or Wave Atmospheric Model (WAM)
would be necessary. 

\subsection{Input/Output utilities} 
The  I/O utilities include generic coupling functionalities that deal with boundary conditions, provide a
restart facility, add surface, lateral and point source fluxes (lower left branch in Fig.~\ref{fig:functionalities}).  

\subsubsection{Generic output}\label{sec:output}

This utility component of MOSSCO provides a generic output facility \texttt{netcdf\_component} for any data that is 
communicated in the coupling framework. The component writes one- to three-dimensional time sliced data into a 
NetCDF \citep[][]{Rew1990} file and adds metadata on the simulation to this output.  
Multiple instances of this component can be used within a simulation, such that output of different variables, differently 
processed data, and output at various output time steps can be recorded. The output component is fully parallellized
with a grid decomposition inherited from one of the coupled science 
components.
In order to reduce interprocess communication during runtime, each write process considers only the part of the data that resides within its compute domain. This comes 
at a cost to the user, who has to postprocess the output tiles to combine for later analysis;  a python script is provided 
with MOSSCO that takes care of joining tiled files. 
 
The generic output also adds meta data that is collected from the system and the user environment when the output is 
written to disk.   Diagnostics about the processing element and run time between output steps are recorded.   The 
structure of the NetCDF output  follows the Climate and Forecast \citep[CF,][]{Eaton2011} convention for physical 
variables, geolocation, units, dimensions and methods modifying variables.  When (mostly biological) terms are not 
available in the controlled vocabulary of CF,  names are built to resemble those contained in the standard.  

\subsubsection{Generic input}\label{sec:input}

The generic  \texttt{netcdf\_input\_component} of MOSSCO reads from NetCDF files and provides the file content 
wrapped in ESMF data structures (fields) to the coupling framework.  It is parallelized identical to the generic output 
component, and inherits its decomposition from other components in the coupled system.  Data can be read for the 
entire domain or for all decomposed compute elements separately.  Upon reading of data, fields can be renamed and 
filtered before they are passed on to the coupled system. 

The input component is typically used to initialize other components, for restarting, to provide boundary conditions, and 
for assimilating data into the coupled system.  The generic input facility supports interpolation of data in time upon 
reading the data, with nearest, most recent, and linear interpolation.   It also supports reading  climatological data and 
translates the climatological timestamp to a simulation present time stamp in the coupling framework.   

\subsection{MOSSCO connectors and mediators}
Information in the form of ESMF states that contain the output fields of every component are communicated to the
ESMF driver; requests for data by every component are also communicated to the ESMF driver component.
MOSSCO connectors are separate components that link output and requested fields between pairwise coupled
components.   MOSSCO informally distinguishes between connector components that do not manipulate the field
data on transfer at all (or only slightly), and mediator components that extract and compute new data out of the input data.

\subsubsection{Link, copy and nudge connectors}\label{sec:linkcopy}

The simplest and default connecting action between components is to share a reference (i.e. a link)
to a single field that resides in memory and can be manipulated by each component; in contrast, the
\texttt{copy\_connector} duplicates a field at coupling time.  The consideration of a link or copy connector is 
important for managing the data flow sequence in a coupled system: the copy mechanism ensures that two coupled  
components work on the same lagged state of data, whereas the link mechanism ensures that each
component works on the most recent data available.

The  \texttt{nudge\_connector} is used to consolidate output from two components by weighted averaging of the
connected fields.  It is typically used as a simple assimiliation tool to drive model states towards observed
states, or to impose boundary conditions.

These connectors can only be applied between components that run on the same grid (but maybe with a different subdomain decomposition).
The \texttt{link\_connector} can only be applied between components with an identical subdomain decomposition
so that the components have access to the same memory.
Components on different grids require regridding, which is currently under development in MOSSCO.

\subsubsection{Transport connector}\label{sec:transport}

A model component qualifies as a transport component when it  offers to
transport  an arbitrary number of tracers in its numerical grid; this facility is
present, for example, in the current \texttt{gotm\_component} and
\texttt{getm\_component}. The \texttt{transport\_connector} provides generic
infrastructure that communicates the tracer fields to be transported to the transporting
component based on the availability of both, the tracer concentrations as well as
their rate of vertical movement independent of the water currents. This connector
is usually called only once per coupled pair of components during the initialization phase.

\subsubsection{Mediators for soil--pelagic coupling}\label{sec:bpcoupler}

One aspect of the generalized coupling infrastructure in MOSSCO is the use of
connecting components that mediate between technically or 
scientifically incompatible data field collections. The soil--pelagic coupling of biogeochemical
model components with a variety of different state variables raises the need for these  mediators.
The use of mediators leaves
the level of data aggregation and dis-aggregation, and unit conversion to the 
coupling routine, instead of requiring specific output from a model component depending
on its coupling partner component.

For soil--pelagic (or benthic--pelagic) coupling, the  \texttt{soil\_pelagic\_connector}
mediates the soil biogeochemistry output towards the pelagic ecosystem input and
the  \texttt{pelagic\_soil\_connector} mediates the pelagic ecosystem output towards
the soil biogeochemistry input, e.g.:
\begin{inparaenum}[(i)]
\item dis-aggregation of dissolved inorganic nitrogen to dissolved ammonium and dissolved nitrate
\item filling missing pelagic state fields for phosphate using the Redfield-equivalent for dissolved inorganic nitrogen
\item calculation of the vertical flux of particulate organic matter (POM) from the water column into the sediment depending 
on POM concentrations in the near-bottom water, its sinking velocity and a sedimentation efficiency depending on the 
near-bottom turbulence. The effective vertical flux is communicated into the pelagic ecosystem component to budget the 
respective loss, and is communicated to the soil biogeochemistry component to account for the respective new mass of 
POM. The mediator also handles
\item dis-aggregation of a single oxygen concentration (allowing positive and negative values) into dissolved oxygen 
concentration, if positive, and dissolved reduced substances, if negative.
\item aggregation of pelagic POM composition (variable nitrogen to carbon ratio) into fixed stoichiometry POM pools in 
the soil biogeochemistry.
\end{inparaenum}

\section{Selected applications as feasibility tests and use cases}\label{sec:apps}

MOSSCO was designed for enhancing flexibility and equitability in environmental data and model coupling. These design goals 
have been helpful in generating new integrated models for coastal research ranging from one-dimensional water-
column to three-dimensional, with applications at different marine stations, transects, and sea domains.  Below, we 
describe from a user perspective the added value and success of the design goals  obtained from using MOSSCO in 
selected applications; here, the focus is not on the scientific outcome of the application (these are described elsewhere 
by, e.g., \citeauthor{Nasermoaddeli2017}, submitted, \citeauthor{Slavik2017}, to be submitted, \citealt{Wirtz2016}, \citealt{Kerimoglu2017}). All setups
described in the use cases are available as open source (with limited forcing data due to space and bandwidth constraints).

\subsection{Helgoland station}\label{sec:helgoland}

The seasonal dynamics of nutrients and turbidity emerges from
the interaction of physical, ecological and biogeochemical processes in the water column and the 
underlying sea floor.  We resolve these dynamics in a coupled application for a 
1D~vertical water column for a station near the German offshore  island Helgoland.  Average water
depth around the island is 25\,m; tidal currents are affected by
the M2 and S2~tides with a characteristic spring--neap cycle, with current velocity not 
exceeding 1 m\,s$^{-1}$. 

The Helgoland 1D application is realized by a coupled system consisting of GOTM hydrodynamics,
the pelagic FABM component with a nutrient--phytoplankton--zooplankton--detritus (NPZD) ecosystem model  \citep{Burchard2005} and two SPM size classes,
interacting with the erosion and sedimentation module,  the sediment component with the OMEXDIA\_P 
early diagenesis submodel, and coupler components for soil--pelagic, pelagic--soil and 
tracer transport. This system and setup is described in more detail by \citet{Hofmeister2014iche}.

\begin{figure*}
\begin{overpic}[grid=false,viewport=0 78 390 330,clip=,width=.8\hsize]{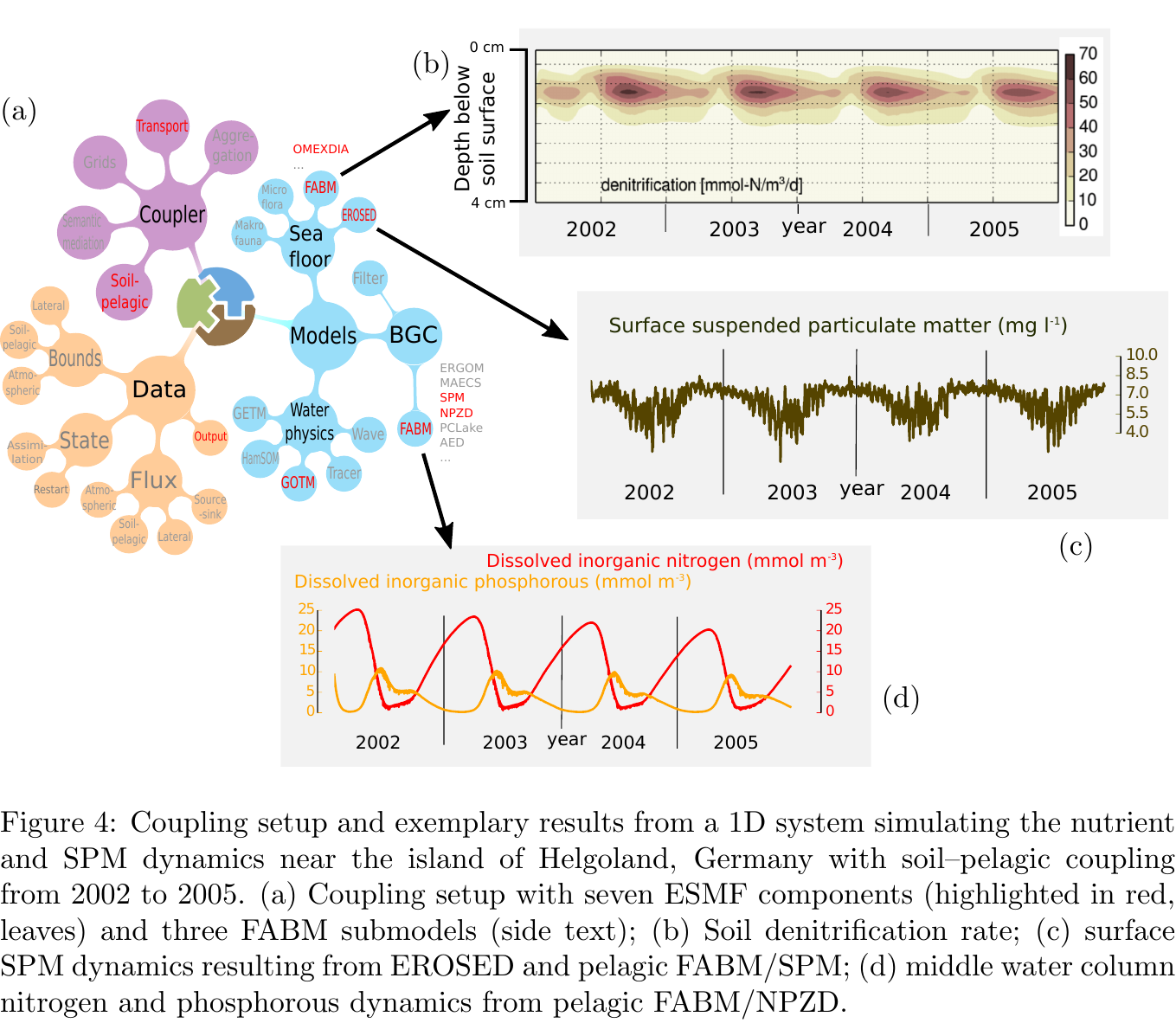}
\end{overpic}
\caption{Coupling setup and exemplary results from a 1D system simulating the nutrient and SPM dynamics near the
island of Helgoland, Germany with soil--pelagic coupling from 2002 to 2005.  (a) Coupling setup with seven ESMF components (highlighted in red, leaves) and three FABM submodels (side text); (b) soil denitrification rate;   (c) surface SPM dynamics resulting from EROSED and pelagic FABM/SPM; (d) middle water column nitrogen and phosphorous dynamics from pelagic FABM/NPZD. }
\label{fig:helgoland}
\end{figure*}

Simulations with this application show a prevailing seasonal cycle in the model states (Fig.~\ref{fig:helgoland}). 
Dissolved nutrients  (referred as dissolved inorganic nitrogen) are taken up by phytoplankton, which fills
the pool of particulate organic nitrogen during the spring bloom (Fig.~\ref{fig:helgoland}d). The particulate organic matter sinks
 into the sediments, where it is remineralized along axis, sub-oxic and anoxic pathways;  denitrification, for example, shows a peak in late summer
 (Fig.~\ref{fig:helgoland}b). At the end of a year, nutrient concentrations are high in the sediment and diffuse back into the
 water column up to winter values of 20--25 mmol\,m$^{-3}$.  The seasonal variation of turbidity is a result of higher
 erosion in winter and reduced vertical transport in summer  (Fig.~\ref{fig:helgoland}c).

\subsection{Idealized coastal 2D transect}\label{sec:transect}
		
The coastal nitrogen cycle is resolved in an idealized coupled system
for a tidal shallow sea. This two-dimensional setup represents a
vertically resolved cross-shore transect of 60\,km length and 5--20\,m
water depth and has  been used  by \citet{Hofmeister2017}
to simulate  sustained horizontal nutrient gradients by particulate matter
transport towards the coast.  Within the MOSSCO coupling framework,
the 2D~transect scenario additionally provides insights into horizontal variability of
erosion/sedimentation and benthic biogeochemistry. 
Its coupling configuration builds on the one used for the 1D~station
Helgoland  (Sect.~\ref{sec:helgoland}); the water-column hydrodynamic model
GOTM, however, is replaced by the 3D~model GETM; a local wave component and
data components for open boundaries and restart has been added.

\begin{figure}
\centering
\begin{overpic}[grid=false,viewport=0 130 365 370,clip=,width=.9\hsize]{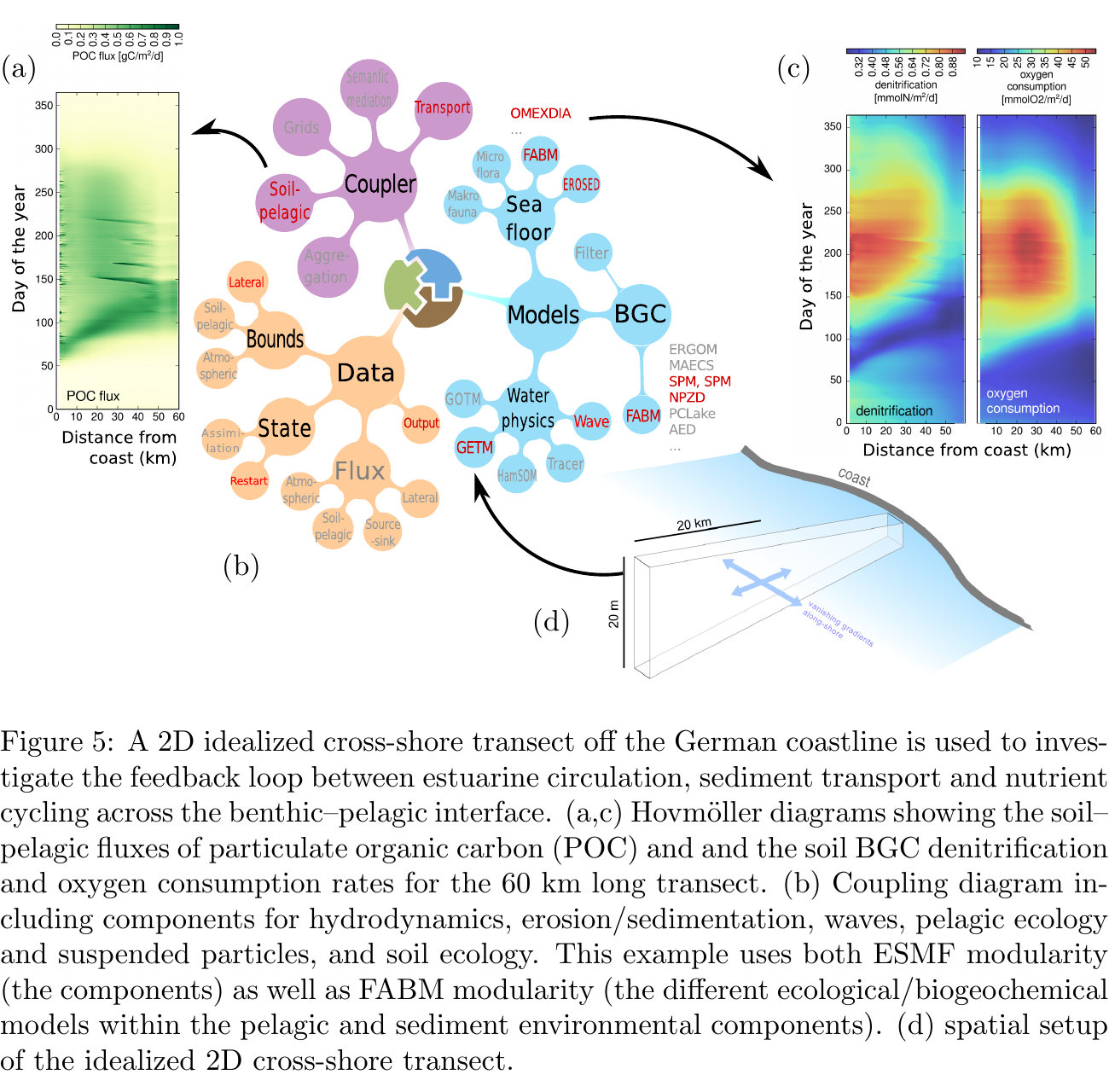}
\end{overpic}
\caption{A 2D idealized cross-shore transect off the German coastline is used to investigate the feedback loop between estuarine
circulation, sediment transport and nutrient cycling across the benthic--pelagic interface.  (a,c) Hovm\"oller diagrams showing the soil--pelagic fluxes of particulate organic carbon (POC) and the soil BGC denitrification and oxygen consumption rates for the 60 km long transect. (b) Coupling diagram including components for hydrodynamics, erosion/sedimentation, waves, pelagic ecology and suspended particles, and soil ecology.  This example uses both ESMF modularity (the components) as well as FABM modularity (the different ecological/biogeochemical models within the pelagic and sediment environmental components).  (d) spatial setup of the idealized 2D cross-shore transect.}
\label{fig:transect}
\end{figure}

Figure \ref{fig:transect} shows exchange fluxes between the water column
and the sediment for one year of simulation.  The simulation of turbidity, as a
result of pelagic SPM transport and resuspension by currents  and wave stress,
provides the light climate for the pelagic ecosystem. The flux of particulate
organic carbon (POC) into the sediment reflects bloom 
events in summer during calm weather conditions. Macrobenthic activity in the sea floor brings fresh organic matter into 
the deeper suboxic layers of the sediment, where denitrification removes nitrogen from the pool of dissolved nutrients. 
The coupled simulation reveals decoupled signals of benthic respiration, denitrification and nutrient reflux into the water 
column, which is not resolved in monolithically coded regional ecosystem models of the North Sea 
\citep{Lorkowski2012,Daewel2013}.

\subsection{Southern North Sea bivalve ecosystem applications}

\begin{figure*}
\centering
\begin{overpic}[grid=false,viewport=0 120 384 460,clip=,width=.9\hsize]{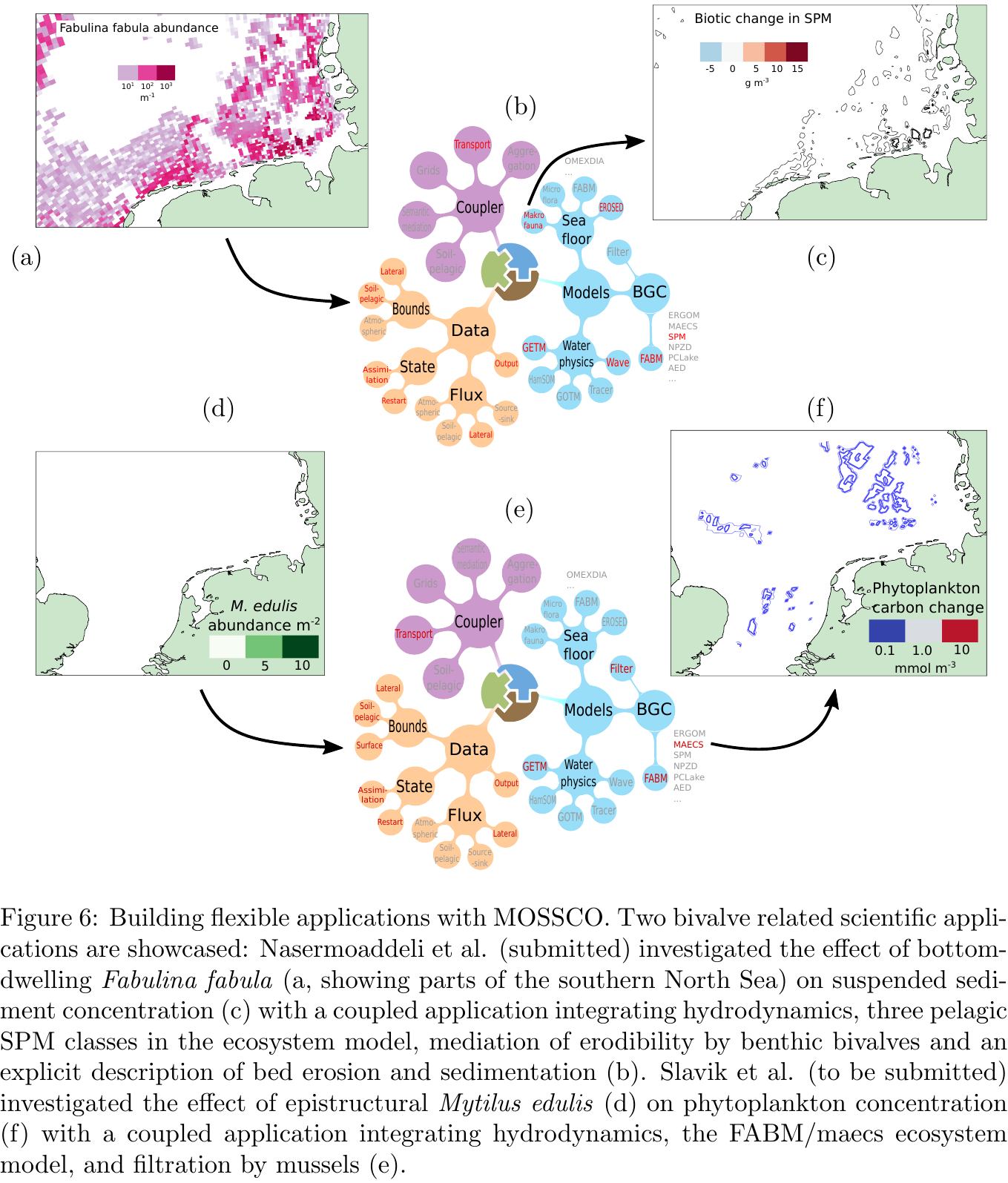}
\end{overpic}
\caption{Building flexible applications with MOSSCO.  Two bivalve related scientific applications are showcased: 
\citeauthor{Nasermoaddeli2017} (submitted) investigated the effect of bottom-dwelling \textit{Fabulina fabula} (a, showing parts of the southern North Sea) on suspended sediment concentration (c) with a coupled application integrating 
hydrodynamics, three pelagic SPM classes in the ecosystem model, mediation of erodibility by benthic bivalves and an 
explicit description of bed erosion and sedimentation (b).  \citeauthor{Slavik2017} (to be submitted) investigated the effect of epistructural \textit{Mytilus edulis} (d) on phytoplankton concentration (f) with a coupled application 
integrating hydrodynamics, the FABM/MAECS ecosystem model, and filtration by mussels (e).}
\label{fig:bivalves}
\end{figure*}

A Southern North Sea (SNS) domain was used in two  studies concerning the
effects of bivalves on the pelagic ecosystem.  \citeauthor{Slavik2017} (to be
submitted) investigated  how the accumulation of epifauna on wind turbine
structures (Fig.~\ref{fig:bivalves}d) impacts pelagic primary production and
ecosystem functioning in the SNS at larger spatial scales. This study is the
first of its kind that extrapolates ecosystem impacts of anthropogenic offshore
wind farm structures from a local to a regional sea scale.  The authors use  a 
MOSSCO coupled system consisting of the hydrodynamic model GETM, the
ecosystem model MAECS as described by \citet{Kerimoglu2017}, the transport
connector, the filter feeder component, and several input and output components
(Fig.~\ref{fig:bivalves}e).   They assess the impact of anthropogenically
enhanced filtration from  blue mussel (\emph{Mytilus edulis}) settlement on
offshore wind farms that are planned to meet the 40-fold increase in offshore
wind electricity in the European Union until~2030; they find a small but 
non-negligible large-scale effect in both phytoplankton stock and primary production, 
which possibly contributes to better water clarity (Fig.~\ref{fig:bivalves}f).

Biological activities of macrofauna on the sea floor mediate suspended sediment
dynamics, at least locally. In the study  by \citeauthor{Nasermoaddeli2017}
(submitted), the large-scale biological contribution of benthic
macrofauna, represented by the key species \emph{Fabulina fabula}
(Fig.~\ref{fig:bivalves}a), to suspension of sediment was investigated. Simulation
results for a typical winter month revealed that SPM is increased not only locally 
but beyond the mussel inhabited zones. This effect is not limited to the near-bed
water layers but can be observed throughout the entire water column, especially
during storm events (Fig.~\ref{fig:bivalves}c).  In this coupled application, the
hydrodynamic model GETM, the pelagic ecosystem component with three SPM
size classes, the erosion--sedimentation and benthic mediation components,
several input and one output components, and the transport connector were
used (Fig.~\ref{fig:bivalves}d).   

\section{Discussion and Outlook}
In merging existing frameworks that address distinct types of modularity and
by developing a super-structure for making the multi-level coupling approach
applicable in coastal research, the MOSSCO system largely  meets the design
goals \emph{flexibility}  and \emph{equitability}. In doing so, structural deficiencies of
legacy models and the need for practical compromises became very
apparent.   

For legacy reasons, \emph{equitability} is the harder to achieve 
design goal.  Both the distribution of compute resources as well as the 
spatial grid definition can be in principle determined by any one of the participating
components; de facto, in marine or aquatic research, they are prescribed by
the hydrodynamic models 
that have so far not been enabled to inherit a grid specification or a resource
distribution from a coupler or coupled system.  With the ongoing development and
diversification of hydrodynamic models, and no immediate benefit for the different physical
models to outsource grid/resource allocation, this situation is not likely to
change.  MOSSCO compromises here by its flexible grid inheritance scheme and
with the grid provisioning component that delivers this information to the 
coupled system whenever a hydrodynamic component is not used.

Beyond grid/resource allocation, however, the \emph{equitability} concept is successfully driving
independent developments of submodules.  We found that indeed experts
in one particular model, e.g. the erosion module, could rely on the 
functionality of the other parts of the system without having to be an expert
themselves in all of the constituent models in the coupled application. 
The limitations to this black-box approach became evident in the scientific
application and evaluation of the coupled model system, which was only
possible when a collaboration with experts in these other model systems was
sought.  By taking away the inaccessibility barrier and by enforcing clear separation
of tasks the modular system stimulated a successful collaboration.  
Sustained granularity also helped to align with ongoing development
in external packages. These can be integrated fast into the coupled system, which
does not rely on specific versions of the externally provided software
unless structural changes occur.
Long-term supported interfaces on the external model side facilitate MOSSCO being
up-to-date with e.g. the fast evolving GETM and FABM code bases.  

When legacy codes were equipped with a framework-agnostic  interface
we encountered  four major difficulties:
\begin{enumerate}
\item For organizing the data flow between the components, MOSSCO
uses standard names and units compatible with the infrastructure and 
library of standard names and units provided in the pelagic component 
for the FABM framework (mostly modelled on CF).  Other components, 
such as the BMIs of wrapped legacy models, do not provide such a 
standard name in their own implementation, and in particular, often do
not adhere to a naming standard. We found ambiguity arising, e.g.,
with temperature to be represented as  \texttt{temperature} vs.\
\texttt{sea\_water\_temperature} vs.\ \texttt{temperature\_in\_water}.
While this can be resolved
based on CF for temperature, most ecological and biogeochemical
quantities currently lack a consistent naming scheme.  The forthcoming
GSN ontology \citep[building on CSDMS names,][]{Peckham2014iemss}
could adequately address this coupling challenge. 

\item Deep subroutine hierarchies of existing models made it difficult
to isolate desired functionality from the structural external overhead.
In one example, where a single functional module was taken out of the 
context of an existing third-party coupled system, the module depended on
many routines dispersed throughout that third-party system repository.

\item Components based on standalone models are developed and tested
with their own I/O infrastructure and typically supply a BMI implementation
only for part of their state and input data fields. A new, coupled application or
data provisioning/request within a coupled system can therefore easily require
a change in the model?s BMI. The implementation potential input and
output for all quantities, including replacement of the entire model-specific I/O
in the BMI is therefore desirable for new developments and refactoring.

\item Two-way coupling of mass and energy fluxes between components has
to be integrated numerically in a conservative way, despite different time
discretization schemes used within the different components. We relied on
conservative integration of the transferred fluxes within pairwise coupled
components for thir respective coupling timestep, which is most flexible for
asynchronous scheduling. The coupled system  itself, however, cannot ensure
the conservative integration of mass and energy fluxes between components. 
Here, the user needs to take care of a correct coupling configuration.
\end{enumerate}

Efforts in making legacy models couplable, either for MOSSCO or 
similar frameworks, however, leads to additional benefits besides the 
immediate applicability in an integrated context. Couplability strictly demands
for communication of sufficient metadata, which stimulates the quality and 
quantity of documentation and of scientific and technical reproducibility of
legacy models. Indeed, transparency has been greatly increased by wrapping legacy
models in the MOSSO context.   All participating components  performed
introspection and leveraging of a collection of metadata at assembly time of the coupled
application and during output.  Transparency is
expected to be continuously increasing by new coupling demands and more
generous metadata provisioning from wrapped science models.  MOSSCO 
is moving towards adopting the Common Information Model~(CIM) that is also 
required by Climate Model Intercomparison Project~(CMIP) participating
coupled models \citep{Eyring2016}.  

With a current small development base of twelve contributors, the
openness concept of MOSSCO in terms of including contributions
from outside the core developer team has not yet been tested; in the
categorization by \citet{DeLaat2007} internal governance with 
simple structure is sufficient at this size.  Formally, external contributions
can be included in MOSSCO by way of contributor license agreements.
The openness concept has been useful  in instigating
discussions about the need for explicit (and preferably open)
licensing of related scientific software and data as demanded in current
open science strategies \citep[e.g.,][]{Scheliga2016}.

So far, scalability in current MOSSCO applications is excellent: 
Strong scaling experiments with a coupled application using GETM,
FABM with MAECS ($\approx$20~additional transported 3D tracers), and FABM 
with OMEXDIA\_P on Jureca \citep{Krause2016} show linear (perfect) speedup
from 100 to 1000~cores, and a small leveling-off (to~85\% of perfect scaling)
at 3000~cores. We have not observed  loss of compute time  due to the
infrastructure and superstructure overhead of ESMF. 

Problems of multi-component systems 
need to be solved in terms of acceptance by the research community. 
Multi-component systems are much harder to be implemented and maintained
by individual groups, where researchers solve coastal ocean problems of a
large range of complexity,  from purely hydrodynamic applications via coupled
hydrodynamic--sediment dynamic applications to fully coupled systems. Many 
academic problems  focus on specific mechanisms and thus do not require
the complete and fully coupled modular system, such that the application of the
full system might mean a large structural overhead and additional workload.
Most researchers would agree on the potential necessity of following a holistic
approach when tackling grand research questions in environmental science
such as related to system responses to anthropogenic intervention. Yet, it is
not clear whether the up-scaling approach inherent to the addition of many
modular components can lead to meaningful results.

As evident form the test cases (\refsec{apps}), MOSSCO also encourages
coupled applications that are far from a complete system level description.  
With few coupled components, the technical threshold to getting an application
running on an arbitrary system is relatively low.  The user can reach a fast first success.
MOSSCO provides a full documentation, step by step recipes, and a public bug tracker;
it adopts abundant error reporting from ESMF and a fail fast design that
stops a coupled applications as soon as a technical error is detected \citep{Shore2004}.
Usability is especially high due to an available master script that compiles,
deploys, and schedules a coupled application. To address a wide range of
users, the system is designed to run on a single processor or on a user's laptop
equally well as on a high-performance computer using several
thousand compute nodes.

An obvious advantage of modular coupling is the opportunity to bridge the gap
between different scientific disciplines. It allows in principle to combine, e.g.,
hydrodynamic models from oceanography with sediment transport models
from coastal engineering. Thus different experts can work on their individual
models but benefit from each others' progress. This seeming advantage, however,
poses also a drawback for modular coupling approaches: An initial effort which is necessary
for individual models to meet the requirements of a modular modelling framework
has to be invested. This will only happen if there is either an urgent pressure
to include specific model capabilities, which will be difficult to include
otherwise, or if convincing examples of possible benefits can be presented.
It cannot be expected that the coastal ocean modelling community
will agree about one coupler or one way of interfacing modules, such that it
will still require considerable implementation work to transfer a module from
one modular system to another. To solve this problem, coupling standards
need to become more general, but in turn this might even increase the
structural overhead in using these systems.

Offsetting these concerns, the separate design of basic and component
interfaces (BMI/CMI) ensures that the effort spent on wrapping an existing
model, or on equipping a new model with a basic model interface is not tied
to a particular coupling framework, or even a particular coupling framework
technology.  A model that  follows BMI principles will be more easily
interfaced to other models no matter what coupler is used.  Wrapped legacy
models from MOSSCO can thus be useful in non-ESMF contexts, as well; 
and models with an existing BMI can be integrated in MOSSCO more easily,
in turn.

One demand for integrative modeling, which is likely best practised in open
and flexible system approaches, arises from
current European Union legislation. The Water Framework Directive and the
Marine Strategic Planning Directive require the description of marine
environmental conditions and the development of action plans to achieve a
good environmental status. These objectives can initially be met by a monitoring
program to determine present-day conditions but ultimately rely on numerical
model studies to evaluate anthropogenic measures. This ecosystem based
approach to management \citep[e.g.,][]{Ruckelshaus2008} demands modelling
systems which are capable of taking into account hydrodynamics, biogeochemistry,
sedimentology and their interactions to properly describe the
environmental status. As further legal requirements can be expected for many
coastal seas worldwide, numerical
modelling systems applied for this task need to be flexible in terms of integrating
additional (e.g., site-specific) processes. In this ongoing process, the initial effort of creating a
modular system may be the only way forward that can take into account all relevant processes
in the long run.

\subsection{Outlook}

The suite of components provided or encapsulated so far meets the demands
that were initially formulated by our users; they already allow for a wide range of
novel coupled applications to investigate the coastal sea.  To stimulate more
collaboration, however, and to bring forward a  general ``ecosystem'' of modular science
components, several legacy models could interface to MOSSCO components
in the near future by building on complementary work at other institutions.  For example,
the Regional Earth
System Model \citep[RegESM,][]{Turuncoglu2013} provides ESMF interfaces for
MITgcm, ROMS  and WAM, amongst others.
Convergence of the development of MOSSCO and RegESM is feasible in the near term.
Also, the recently developed Icosahedral Non-Hydrostatic Atmospheric Model
\citep[ICON][]{Zaengli2015} is currently being equipped with an ESMF component model
interface.

Once ESMF interfaces have been developed for a legacy model, it is desirable that these
developments move out of the coupler system and are integrated into the development of
the legacy model itself.  This has been successfully  achieved with the ESMF interface
for the hydrodynamic model GETM, which is now distributed with the GETM code.
Much of the utility layer  developed in MOSSCO, or likewise in MAPL or
in the ESMF extension of the WRF model, are expected to be propagated upstream into
the framework ESMF itself.   

The interoperability of current coupling standards will increase.   While currently there are
three flavors of ESMF: basic  ESMF as in MOSSCO, ESMF/MAPL as in the GEOS-5 system,
or ESMF/NUOPC  as in the RegESM,  only a minor effort would be required to provide the basic
ESMF and ESMF/MAPL implementations with a NUOPC cap and make them interoperable with
the entire ESMF ecosystem.    Even a coupling of ESMF based systems to OASIS/MCT based
systems has been proposed; and investigation is ongoing on a 
coupling of MOSSCO to the formal BMI for CSDMS.  

\conclusions

We problematized both the primacy of hydrodynamic models and the limited modularity in
coupled coastal modeling that can stand in the way of developing and applying novel and
diverse biogeochemical process descriptions. Such developmental potential is likely
needed to progress towards holistic regional coastal systems models.  We presented the
novel Modular System for Shelves and Coasts (MOSSCO), that is  built on coupling
concepts centered around equitability and flexibility to resolve the issue of obstructed modularity.
These concepts bring about also openness, usability, transparency and scalability. MOSSCO
as an actual Fortran implementation of this concept includes the wrapped Framework
for Aquatic Biogeochemical Models (FABM) and a usability layer for the Earth System Modeling
Framework (ESMF).

MOSSCO's design principles emphasize basic couplability and rich meta information.  Basic
couplability requires that models communicate about flow control, compute resources, and
about exchanged data and metadata. We demonstrated that  the design principles flexibility
and equitability enable the building of complex coupled models that adequately
represent the complexity found in environmental modelling. In this first version, the MOSSCO
software wrapped several existing legacy models with basic model interfaces (BMI); 
we added ESMF-specific component model interfaces (CMI) to these wrappers and other
models and frameworks to build a suite of ESMF components that when coupled represent
a small part of a holistic coastal system.  These components deal with hydrodynamics,
waves, pelagic and sediment ecology and biogeochemistry, river loads, erosion,
resuspension, biotic sediment modification and filter feeding. 

In selected applications, each with a different research question, the applicability of the coupled
system was successfully tested. MOSSCO  facilitates the development of new coupled
applications for studying coastal processes that extend from the atmosphere through
the water column into the sea bed, and that range from laboratory studies to 3D
simulation studies of a regional sea. This system meets an infrastructural need that is
defined by experimenters and process modellers who develop biogeochemical, physical,
sedimentological or ecological models at the lab scale first and who would like to
seamlessly embed these models into the complex coupled three-dimensional coastal
system. This upscaling procedure may ultimately support also the global Earth System community. 

\codedataavailability{The MOSSCO software is licensed under the GNU General Public
License~3.0, a copyleft open  source license that allows the  use, distribution and
modification of the software under the same terms.  All documentation for MOSSCO is
licensed under the Creative Commons Attribution Share-Alike~4.0 (CC-by-SA), a 
copyleft open document license that allows use, distribution and modification of the
documentation under the same terms. 

Development code and documentation are currently  primarily hosted on Sourceforge
(\href{https://sf.net/p/mossco/code}{https://sf.net/p/mossco/code}) and mirrored on Github
(\href{https://github.com/platipodium/mossco-code}{https://github.com/platipodium/mossco-code}).
The release version 1.0.1 is permanently archived
on Zenodo and accessible  under the digital object identifier
\href{http://dx.doi.org/10.5281/zenodo.438922}{doi:10.5281/zenodo.438922}.  
All wrapped legacy models are open source and freely available from the developing institutions;  
free registration is required  for accessing the Delft3D system at Deltares.
Selected test cases are available from a separate Sourceforge repository
\href{https://sf.net/p/mossco/setups}{https://sf.net/p/mossco/setups}, where all of the data on
which the presented use cases are based are freely available, with the 
exception of the meteorological forcing fields.  These are, for example, available at request online
at \href{http://www.coastdat.de}{http://www.coastdat.de}, from the coastDat model based
data base developed for the assessment of long-term changes by Helmholtz-Zentrum Geesthacht \citep{Geyer2014}.
}\label{sec:availability}

\authorcontribution{C.L., R.H., K.K., H.N developed the MOSSCO components (CMI) and wrappers (BMI).  K.W, C.L., and K.K.
designed the coupling philosophy, C.L. developed the user interface and the utility library. K.W., H.N., R.H., O.K. and C.L. 
carried out and analysed simulations, based on contributions from all authors.  C.L., K.W. and R.H. wrote the manuscript with contributions from all other authors.}

\competinginterests{The authors declare that they have no conflict of interest.}

\begin{acknowledgements}
MOSSCO is a project funded under the K\"ustenforschung Nordsee--Ostsee programme of the Forschung f\"ur 
Nachhaltigkeit (FONA) agenda of the German Ministry of Education and Science (BMBF) under grant agreements 
03F0667A, 03F0667B, and 03FO668A.  This research contributes to the PACES~II programme of the Hermann von 
Helmholtz-Gemeinschaft Deutscher Forschungszentren.
Further financial support for K.K.\ and H.B.\ was provided by the Collaborative Research Centre TRR181
on Energy Transfers in Atmosphere and Ocean funded by the German Research Foundation (DFG).
We thank those MOSSCO developers that are not co-authors 
of this paper, amongst them Markus Kreus, Ulrich K\"orner and Niels Weiher, and acknowledge the support of 
Wenyan Zhang in preparing the model setups.  This research is based on tremendous efforts by the open 
source community, including but not limited to the developers of Delft3D, GETM, GOTM, FABM, ESMF, OpenMPI, 
Python, GCC and NetCDF who share their codes openly.
\end{acknowledgements}

\bibliographystyle{copernicus}
\bibliography{Lemmen2017_etal_geoscientificmodeldevelopment}

\begin{thebibliography}{67}
\providecommand{\natexlab}[1]{#1}
\providecommand{\url}[1]{{\tt #1}}
\providecommand{\urlprefix}{URL }
\expandafter\ifx\csname urlstyle\endcsname\relax
  \providecommand{\doi}[1]{doi:\discretionary{}{}{}#1}\else
  \providecommand{\doi}{doi:\discretionary{}{}{}\begingroup
  \urlstyle{rm}\Url}\fi

\bibitem[{Alexander and Easterbrook(2015)}]{Alexander2015}
Alexander, K. and Easterbrook, S.~M.: {The software architecture of climate
  models: a graphical comparison of CMIP5 and EMICAR5 configurations},
  Geoscientific Model Development, 8, 1221--1232,
  \doi{10.5194/gmd-8-1221-2015}, 2015.

\bibitem[{Aumont et~al.(2015)Aumont, Eth{\'{e}}, Tagliabue, Bopp, and
  Gehlen}]{Aumont2015}
Aumont, O., Eth{\'{e}}, C., Tagliabue, A., Bopp, L., and Gehlen, M.:
  {PISCES-v2: an ocean biogeochemical model for carbon and ecosystem studies},
  Geoscientific Model Development, 8, 2465--2513,
  \doi{10.5194/gmd-8-2465-2015}, 2015.

\bibitem[{Azhikodan and Yokoyama(2016)}]{Azhikodan2016}
Azhikodan, G. and Yokoyama, K.: {Spatio-temporal variability of phytoplankton
  (Chlorophyll-a) in relation to salinity, suspended sediment concentration,
  and light intensity in a macrotidal estuary}, Continental Shelf Research,
  126, 15--26, \doi{10.1016/j.csr.2016.07.006}, 2016.

\bibitem[{Balaji et~al.(2007)Balaji, Adcroft, and Liang}]{Balaji2007}
Balaji, V., Adcroft, A., and Liang, Z.: {Gridspec: A standard for the
  description of grids used in Earth System models}, Tech. rep., National
  Oceanographic and Atmospheric Administration, Princeton, NJ, 2007.

\bibitem[{Balaji et~al.(2016)Balaji, Benson, Wyman, and Held}]{Balaji2016}
Balaji, V., Benson, R., Wyman, B., and Held, I.: {Coarse-grained component
  concurrency in Earth system modeling: parallelizing atmospheric radiative
  transfer in the GFDL AM3 model using the Flexible Modeling System coupling
  framework}, Geoscientific Model Development, 9, 3605--3616,
  \doi{10.5194/gmd-9-3605-2016}, 2016.

\bibitem[{Bayne et~al.(1993)Bayne, Iglesias, and Hawkins}]{Bayne1993}
Bayne, B.~L., Iglesias, J., and Hawkins, A. J.~S.: {Feeding behaviour of the
  mussel, Mytilus edulis: responses to variations in quantity and organic
  content of the seston}, Journal of the Marine Biological Association of the
  United Kingdom, 73, 813--829, 1993.

\bibitem[{Breugem and Holthuijsen(2007)}]{Breugem2007}
Breugem, W.~a. and Holthuijsen, L.~H.: {Generalized Shallow Water Wave Growth
  from Lake George}, Journal of Waterway, Port, Coastal, and Ocean Engineering,
  133, 173--182, \doi{10.1061/(ASCE)0733-950X(2007)133:3(173)}, 2007.

\bibitem[{Bruggeman and Bolding(2014)}]{Bruggeman2014}
Bruggeman, J. and Bolding, K.: {A general framework for aquatic biogeochemical
  models}, Environmental Modelling and Software, 61, 249--265,
  \doi{10.1016/j.envsoft.2014.04.002}, 2014.

\bibitem[{Burchard and Bolding(2002)}]{Burchard2002ec}
Burchard, H. and Bolding, K.: {GETM -- a General Estuarine Transport Model.
  Scientific Documentation}, Tech. Rep. EUR 20253 EN, European Commission,
  2002.

\bibitem[{Burchard et~al.(1999)Burchard, Bolding, and
  Villarreal}]{Burchard1999}
Burchard, H., Bolding, K., and Villarreal, M.~R.: {GOTM -- a General Ocean
  Turbulence Model. Theory, implementation and test cases}, Tech. Rep. EUR
  18745 EN, European Commission, 1999.

\bibitem[{Burchard et~al.(2005)Burchard, Deleersnijder, and
  Meister}]{Burchard2005}
Burchard, H., Deleersnijder, E., and Meister, A.: {Application of modified
  Patankar schemes to stiff biogeochemical models for the water column}, Ocean
  Dynamics, 55, 326--337, \doi{10.1007/s10236-005-0001-x}, 2005.

\bibitem[{Burchard et~al.(2006)Burchard, Bolding, K{\"{u}}hn, Meister, Neumann,
  and Umlauf}]{Burchard2006jms}
Burchard, H., Bolding, K., K{\"{u}}hn, W., Meister, A., Neumann, T., and
  Umlauf, L.: {Description of a flexible and extendable
  physical–biogeochemical model system for the water column}, Journal of
  Marine Systems, 61, 180--211, \doi{10.1016/j.jmarsys.2005.04.011}, 2006.

\bibitem[{Butensch{\"{o}}n et~al.(2016)Butensch{\"{o}}n, Clark, Aldridge,
  Allen, Artioli, Blackford, Bruggeman, Cazenave, Ciavatta, Kay, Lessin, van
  Leeuwen, van~der Molen, de~Mora, Polimene, Sailley, Stephens, and
  Torres}]{Butenschoen2016}
Butensch{\"{o}}n, M., Clark, J., Aldridge, J.~N., Allen, J.~I., Artioli, Y.,
  Blackford, J., Bruggeman, J., Cazenave, P.~W., Ciavatta, S., Kay, S., Lessin,
  G., van Leeuwen, S., van~der Molen, J., de~Mora, L., Polimene, L., Sailley,
  S., Stephens, N., and Torres, R.: {ERSEM 15.06: a generic model for marine
  biogeochemistry and the ecosystem dynamics of the lower trophic levels},
  Geoscientific Model Development, 9, 1293--1339,
  \doi{10.5194/gmd-9-1293-2016}, 2016.

\bibitem[{Cazenave et~al.(2016)Cazenave, Torres, and Allen}]{Cazenave2016}
Cazenave, P.~W., Torres, R., and Allen, J.~I.: {Unstructured grid modelling of
  offshore wind farm impacts on seasonally stratified shelf seas}, Progress in
  Oceanography, 145, 25--41, \doi{10.1016/j.pocean.2016.04.004}, 2016.

\bibitem[{Cossarini et~al.(2017)Cossarini, Querin, Solidoro, Sannino, Lazzari,
  {Di Biagio}, and Bolzon}]{Cossarini2017}
Cossarini, G., Querin, S., Solidoro, C., Sannino, G., Lazzari, P., {Di Biagio},
  V., and Bolzon, G.: {Development of BFMCOUPLER (v1.0), the coupling scheme
  that links the MITgcm and BFM models for ocean biogeochemistry simulations},
  Geoscientific Model Development, 10, 1423--1445,
  \doi{10.5194/gmd-10-1423-2017}, 2017.

\bibitem[{Daewel and Schrum(2013)}]{Daewel2013}
Daewel, U. and Schrum, C.: {Simulating long-term dynamics of the coupled North
  Sea and Baltic Sea ecosystem with ECOSMO II: Model description and
  validation}, Journal of Marine Systems, 119-120, 30--49,
  \doi{10.1016/j.jmarsys.2013.03.008}, 2013.

\bibitem[{de~Deckere et~al.(2001)de~Deckere, Tolhurst, and
  de~Brouwer}]{DeDeckere2001}
de~Deckere, E. M. G.~T., Tolhurst, T.~J., and de~Brouwer, J. F.~C.:
  {Destabilization of cohesive intertidal sediments by infauna}, Estuarine,
  Coastal and Shelf Science, 53, 665--669, \doi{10.1006/ecss.2001.0811}, 2001.

\bibitem[{de~Laat(2007)}]{DeLaat2007}
de~Laat, P.~B.: {Governance of open source software: state of the art}, Journal
  of Management {\&} Governance, 11, 165--177, \doi{10.1007/s10997-007-9022-9},
  2007.

\bibitem[{Dunne et~al.(2012)Dunne, John, Adcroft, Griffies, Hallberg,
  Shevliakova, Stouffer, Cooke, Dunne, Harrison, Krasting, Malyshev, Milly,
  Phillipps, Sentman, Samuels, Spelman, Winton, Wittenberg, and
  Zadeh}]{Dunne2012}
Dunne, J.~P., John, J.~G., Adcroft, A.~J., Griffies, S.~M., Hallberg, R.~W.,
  Shevliakova, E., Stouffer, R.~J., Cooke, W., Dunne, K.~A., Harrison, M.~J.,
  Krasting, J.~P., Malyshev, S.~L., Milly, P. C.~D., Phillipps, P.~J., Sentman,
  L.~T., Samuels, B.~L., Spelman, M.~J., Winton, M., Wittenberg, A.~T., and
  Zadeh, N.: {GFDL's ESM2 Global Coupled Climate–Carbon Earth System Models.
  Part I: Physical Formulation and Baseline Simulation Characteristics},
  Journal of Climate, 25, 6646--6665, \doi{10.1175/JCLI-D-11-00560.1}, 2012.

\bibitem[{Eaton et~al.(2011)Eaton, Gregory, Centre, Office, Drach, Taylor,
  Hankin, Caron, and Signell}]{Eaton2011}
Eaton, B., Gregory, J., Centre, H., Office, U. K.~M., Drach, B., Taylor, K.,
  Hankin, S., Caron, J., and Signell, R.: {NetCDF Climate and Forecast (CF)
  Metadata Conventions}, Tech. rep., cfconventions.org, 2011.

\bibitem[{{ESMF Joint Specification Team}(2013)}]{ESMF2013usr}
{ESMF Joint Specification Team}: {Earth System Modeling Framework User Guide
  Version 6.3.0}, Tech. rep., National Oceanic and Atmospheric Administration,
  Boulder, CO, 2013.

\bibitem[{Eyring et~al.(2016)Eyring, Bony, Meehl, Senior, Stevens, Stouffer,
  and Taylor}]{Eyring2016}
Eyring, V., Bony, S., Meehl, G.~A., Senior, C.~A., Stevens, B., Stouffer,
  R.~J., and Taylor, K.~E.: {Overview of the Coupled Model Intercomparison
  Project Phase 6 (CMIP6) experimental design and organization}, Geoscientific
  Model Development, 9, 1937--1958, \doi{10.5194/gmd-9-1937-2016}, 2016.

\bibitem[{Geyer(2014)}]{Geyer2014}
Geyer, B.: {High-resolution atmospheric reconstruction for Europe 1948–2012:
  coastDat2}, Earth System Science Data, 6, 147--164,
  \doi{10.5194/essd-6-147-2014}, 2014.

\bibitem[{Gr{\"{a}}we et~al.(2015)Gr{\"{a}}we, Holtermann, Klingbeil, and
  Burchard}]{Graewe2015}
Gr{\"{a}}we, U., Holtermann, P., Klingbeil, K., and Burchard, H.: {Advantages
  of vertically adaptive coordinates in numerical models of stratified shelf
  seas}, Ocean Modelling, 92, 56--68, \doi{10.1016/j.ocemod.2015.05.008}, 2015.

\bibitem[{Harms(1997)}]{Harms1997}
Harms, I.: {Water mass transformation in the Barents Sea — application of the
  Hamburg Shelf Ocean Model (HamSOM)}, ICES Journal of Marine Science, 54,
  351--365, \doi{10.1006/jmsc.1997.0226}, 1997.

\bibitem[{Hill et~al.(2004)Hill, DeLuca, Balaji, Suarez, and {Da
  Silva}}]{Hill2004}
Hill, C., DeLuca, C., Balaji, V., Suarez, M., and {Da Silva}, A.: {The
  architecture of the Earth system modeling framework}, Computing in Science
  and Engineering, 6, 18--28, \doi{10.1109/MCISE.2004.1255817}, 2004.

\bibitem[{Hinners et~al.(2015)Hinners, Hofmeister, and Hense}]{Hinners2015}
Hinners, J., Hofmeister, R., and Hense, I.: {Modeling the Role of pH on Baltic
  Sea Cyanobacteria}, Life, 5, 1204--1217, \doi{10.3390/life5021204}, 2015.

\bibitem[{Hofmeister et~al.(2010)Hofmeister, Burchard, and
  Beckers}]{Hofmeister2010}
Hofmeister, R., Burchard, H., and Beckers, J.-M.: {Non-uniform adaptive
  vertical grids for 3D numerical ocean models}, Ocean Modelling, 33, 70--86,
  2010.

\bibitem[{Hofmeister et~al.(2014)Hofmeister, Lemmen, Kerimoglu, Wirtz, and
  Nasermoaddeli}]{Hofmeister2014iche}
Hofmeister, R., Lemmen, C., Kerimoglu, O., Wirtz, K.~W., and Nasermoaddeli,
  M.~H.: {The predominant processes controlling vertical nutrient and suspended
  matter fluxes across domains - using the new MOSSCO system form coastal sea
  sediments up to the atmosphere}, in: 11th International Conference on
  Hydroscience and Engineering, edited by Lehfeldt, R. and Kopmann, vol.~28,
  Hamburg, Germany, 2014.

\bibitem[{Hofmeister et~al.(2017)Hofmeister, Fl{\"{o}}ser, and
  Schartau}]{Hofmeister2017}
Hofmeister, R., Fl{\"{o}}ser, G., and Schartau, M.: {Estuary-type circulation
  as a factor sustaining horizontal nutrient gradients in freshwater-influenced
  coastal systems}, Geo-Marine Letters, 37, 179--192,
  \doi{10.1007/s00367-016-0469-z}, 2017.

\bibitem[{Hu et~al.(2016)Hu, Bolding, Bruggeman, Jeppesen, Flindt, van Gerven,
  Janse, Janssen, Kuiper, Mooij, and Trolle}]{Hu2016}
Hu, F., Bolding, K., Bruggeman, J., Jeppesen, E., Flindt, M.~R., van Gerven,
  L., Janse, J.~H., Janssen, A. B.~G., Kuiper, J.~J., Mooij, W.~M., and Trolle,
  D.: {FABM-PCLake – linking aquatic ecology with hydrodynamics},
  Geoscientific Model Development, 9, 2271--2278,
  \doi{10.5194/gmd-9-2271-2016}, 2016.

\bibitem[{Jagers(2010)}]{Jagers2010}
Jagers, H. R. A.~B.: {Linking Data , Models and Tools : An Overview}, in:
  International Environmental Modelling and Software, edited by {David A.
  Swayne, Wanhong Yang, A. A. Voinov, A. Rizzoli}, T.~F., Ottawa, 2010.

\bibitem[{J{\"{o}}ckel et~al.(2005)J{\"{o}}ckel, Sander, Kerkweg, Tost, and
  Lelieveld}]{Joeckel2005}
J{\"{o}}ckel, P., Sander, R., Kerkweg, A., Tost, H., and Lelieveld, J.:
  {Technical Note: The Modular Earth Submodel System (MESSy) - a new approach
  towards Earth System Modeling}, Atmospheric Chemistry and Physics, 5,
  433--444, \doi{10.5194/acp-5-433-2005}, 2005.

\bibitem[{Jones(1999)}]{Jones1999mwr}
Jones, P.~W.: {First- and Second-Order Conservative Remapping Schemes for Grids
  in Spherical Coordinates}, Monthly Weather Review, 127, 2204--2210,
  \doi{10.1175/1520-0493(1999)127<2204:FASOCR>2.0.CO;2}, 1999.

\bibitem[{Kerimoglu et~al.(2017)Kerimoglu, Hofmeister, Maerz, and {Wenzel
  Wirtz}}]{Kerimoglu2017}
Kerimoglu, O., Hofmeister, R., Maerz, J., and {Wenzel Wirtz}, K.: {A novel
  acclimative biogeochemical model and its implementation to the southern North
  Sea}, Biogeosciences Discussions, p.~33, \doi{10.5194/bg-2017-104}, 2017.

\bibitem[{Klingbeil and Burchard(2013)}]{Klingbeil2013}
Klingbeil, K. and Burchard, H.: {Implementation of a direct nonhydrostatic
  pressure gradient discretisation into a layered ocean model}, Ocean
  Modelling, 65, 64--77, \doi{10.1016/j.ocemod.2013.02.002}, 2013.

\bibitem[{Krause and Th{\"{o}}rnig(2016)}]{Krause2016}
Krause, D. and Th{\"{o}}rnig, P.: {JURECA: General-purpose supercomputer at
  J{\"{u}}lich Supercomputing Centre}, Journal of large-scale research
  facilities JLSRF, 2, A62, \doi{10.17815/jlsrf-2-121}, 2016.

\bibitem[{Lorkowski et~al.(2012)Lorkowski, P{\"{a}}tsch, Moll, and
  K{\"{u}}hn}]{Lorkowski2012}
Lorkowski, I., P{\"{a}}tsch, J., Moll, A., and K{\"{u}}hn, W.: {Estuarine ,
  Coastal and Shelf Science Interannual variability of carbon fluxes in the
  North Sea from 1970 to 2006 e Competing effects of abiotic and biotic drivers
  on the gas-exchange of CO 2}, Estuarine, Coastal and Shelf Science, 100,
  38--57, \doi{10.1016/j.ecss.2011.11.037}, 2012.

\bibitem[{Lovelock and Margulis(1974)}]{Lovelock1974}
Lovelock, J.~E. and Margulis, L.: {Atmospheric homeostasis by and for the
  biosphere: the gaia hypothesis}, Tellus, 26, 2--10,
  \doi{10.1111/j.2153-3490.1974.tb01946.x}, 1974.

\bibitem[{Maerz et~al.(2011)Maerz, Verney, Wirtz, and Feudel}]{Maerz2011}
Maerz, J., Verney, R., Wirtz, K.~W., and Feudel, U.: {Modeling flocculation
  processes: Intercomparison of a size class-based model and a
  distribution-based model}, Continental Shelf Research, 31, S84--S93,
  \doi{10.1016/j.csr.2010.05.011}, 2011.

\bibitem[{Manabe(1969)}]{Manabe1969}
Manabe, S.: {Climate and the ocean circulation II. The atmospheric circulation
  and the effect of heat transfer by ocean currents}, Monthly Weather Review,
  97, 775--805, 1969.

\bibitem[{Margalef(1963)}]{Margalef1963}
Margalef, R.: {On Certain Unifying Principles in Ecology}, The American
  Naturalist, 97, 357--374, \doi{10.1086/282286}, 1963.

\bibitem[{Moghimi et~al.(2013)Moghimi, Klingbeil, Gr{\"{a}}we, and
  Burchard}]{Moghimi2013a}
Moghimi, S., Klingbeil, K., Gr{\"{a}}we, U., and Burchard, H.: {A direct
  comparison of a depth-dependent Radiation stress formulation and a Vortex
  force formulation within a three-dimensional coastal ocean model}, Ocean
  Modelling, 70, 132--144, \doi{10.1016/j.ocemod.2012.10.002}, 2013.

\bibitem[{Nasermoaddeli et~al.(2014)Nasermoaddeli, Lemmen, Hofmeister,
  K{\"{o}}sters, and Klingbeil}]{Nasermoaddeli2014ich}
Nasermoaddeli, M.~H., Lemmen, C., Hofmeister, R., K{\"{o}}sters, F., and
  Klingbeil, K.: {The Benthic Geoecology Model within the Modular System for
  Shelves and Coasts (MOSSCO)}, in: 11th International Conference on
  Hydroinformatics, 2014.

\bibitem[{Nasermoaddeli et~al.(2017)Nasermoaddeli, Lemmen, K{\"{o}}sters,
  Kerimoglu, Hofmeister, Klingbeil, Burchard, and Wirtz}]{Nasermoaddeli2017}
Nasermoaddeli, M.~H., Lemmen, C., K{\"{o}}sters, F., Kerimoglu, O., Hofmeister,
  R., Klingbeil, K., Burchard, H., and Wirtz, K.~W.: {Large scale effect of
  macrofauna Tellina fabula on suspended sediment transport in the southern
  Nord See}, Estuarine, Coastal and Shelf Science, submitted, 2017.

\bibitem[{Partheniades(1965)}]{Partheniades1965}
Partheniades, E.: {Erosion and Deposition of Cohesive Soils}, Journal of the
  Hydraulics Division, 91, 105--139, 1965.

\bibitem[{Peckham(2014)}]{Peckham2014iemss}
Peckham, S.~D.: {The CSDMS Standard Names: Cross-Domain Naming Conventions for
  Describing Process Models, Data Sets and Their Associated Variables}, in:
  International Environmental Modelling and Software Society (iEMSs), 2014.

\bibitem[{Peckham et~al.(2013)Peckham, Hutton, and Norris}]{Peckham2013}
Peckham, S.~D., Hutton, E.~W., and Norris, B.: {A component-based approach to
  integrated modeling in the geosciences: The design of CSDMS}, Computers {\&}
  Geosciences, 53, 3--12, \doi{10.1016/j.cageo.2012.04.002}, 2013.

\bibitem[{Rew and Davis(1990)}]{Rew1990}
Rew, R. and Davis, G.: {NetCDF: An Interface for Scientific Data Access}, IEEE
  Computer Graphics and Applications, 10, 76--82, \doi{10.1109/38.56302}, 1990.

\bibitem[{Ruckelshaus et~al.(2008)Ruckelshaus, Klinger, Knowlton, and
  DeMASTER}]{Ruckelshaus2008}
Ruckelshaus, M., Klinger, T., Knowlton, N., and DeMASTER, D.~P.: {Marine
  Ecosystem-based Management in Practice: Scientific and Governance
  Challenges}, BioScience, 58, 53, \doi{10.1641/B580110}, 2008.

\bibitem[{Scheliga et~al.(2016)Scheliga, Pampel, Bernstein, Bruch, zu~Castell,
  Diesmann, Fritzsch, Fuhrmann, Haas, Hammitzsch, {M., L{\"{a}}hnemann},
  McHardy, Konrad, Scharnberg, Schreiber, and Steglich}]{Scheliga2016}
Scheliga, K.~S., Pampel, H., Bernstein, E., Bruch, C., zu~Castell, W.,
  Diesmann, M., Fritzsch, B., Fuhrmann, J., Haas, H., Hammitzsch, {M.,
  L{\"{a}}hnemann}, D., McHardy, A., Konrad, U., Scharnberg, G., Schreiber, A.,
  and Steglich, D.: {Helmholtz Open Science Workshop „Zugang zu und
  Nachnutzung von wissenschaftlicher Software“ {\#}hgfos16}, Tech. rep.,
  Deutsches GeoForschungsZentrum GFZ, Potsdam, \doi{10.2312/lis.17.01}, 2016.

\bibitem[{Shang et~al.(2014)Shang, Fang, Zhao, He, and Cui}]{Shang2014}
Shang, Q.~Q., Fang, H.~W., Zhao, H.~M., He, G.~J., and Cui, Z.~H.: {Biofilm
  effects on size gradation, drag coefficient and settling velocity of sediment
  particles}, International Journal of Sediment Research, 29, 471--480,
  \doi{10.1016/S1001-6279(14)60060-3}, 2014.

\bibitem[{Shore(2004)}]{Shore2004}
Shore, J.: {Fail fast}, IEEE Software, 21, 21--25,
  \doi{10.1109/MS.2004.1331296}, 2004.

\bibitem[{Slavik et~al.(2017)Slavik, Lemmen, Zhang, Kerimoglu, and
  Wirtz}]{Slavik2017}
Slavik, K., Lemmen, C., Zhang, W., Kerimoglu, O., and Wirtz, K.~W.: {The large
  scale impact of offshore windfarm structures on pelagic primary production in
  the southern North Sea}, Hydrobiologia, submitted, 2017.

\bibitem[{Soetaert et~al.(1996)Soetaert, Herman, and Middelburg}]{Soetaert1996}
Soetaert, K., Herman, P. M.~J., and Middelburg, J.~J.: {Dynamic response of
  deep-sea sediments to seasonal variations: A model}, Limnology and
  Oceanography, 41, 1651--1668, 1996.

\bibitem[{Suarez et~al.(2007)Suarez, Trayanov, da~Silva, Hill, and
  Schopf}]{Suarez2007}
Suarez, M., Trayanov, A., da~Silva, A., Hill, C., and Schopf, P.: {An
  Introduction to MAPL}, Tech. rep., Goddard Fluid Dynamics Laboratory,
  Princeton, NJ, 2007.

\bibitem[{Tansley(1935)}]{Tansley1935}
Tansley, A.~G.: {The Use and Abuse of Vegetational Concepts and Terms},
  Ecology, 16, 284--307, \doi{10.2307/1930070}, 1935.

\bibitem[{Turuncoglu et~al.(2013)Turuncoglu, Dalfes, Murphy, and
  DeLuca}]{Turuncoglu2013}
Turuncoglu, U.~U., Dalfes, N., Murphy, S., and DeLuca, C.: {Toward
  self-describing and workflow integrated Earth system models: A coupled
  atmosphere-ocean modeling system application}, Environmental Modelling and
  Software, 39, 247--262, \doi{10.1016/j.envsoft.2012.02.013}, 2013.

\bibitem[{Valcke(2013)}]{Valcke2013}
Valcke, S.: {The OASIS3 coupler: a European climate modelling community
  software}, Geoscientific Model Development, 6, 373--388,
  \doi{10.5194/gmd-6-373-2013}, 2013.

\bibitem[{{Van Pham} et~al.(2014){Van Pham}, Brauch, Dieterich, Frueh, and
  Ahrens}]{VanPham2014}
{Van Pham}, T., Brauch, J., Dieterich, C., Frueh, B., and Ahrens, B.: {New
  coupled atmosphere-ocean-ice system COSMO-CLM/NEMO: assessing air temperature
  sensitivity over the North and Baltic Seas}, Oceanologia, 56, 167--189,
  \doi{10.5697/oc.56-2.167}, 2014.

\bibitem[{van Rijn(2007)}]{VanRijn2007}
van Rijn, L.~C.: {Unified View of Sediment Transport by Currents and Waves. II:
  Suspended Transport}, Journal of Hydraulic Engineering, 133, 668--689,
  \doi{10.1061/(ASCE)0733-9429(2007)133:6(668)}, 2007.

\bibitem[{Vernadsky(1998)}]{Vernadsky1998}
Vernadsky, V.~I.: {The Biosphere}, Springer, 1998.

\bibitem[{Warner et~al.(2008)Warner, Perlin, and Skyllingstad}]{Warner2008ems}
Warner, J.~C., Perlin, N., and Skyllingstad, E.~D.: {Using the Model Coupling
  Toolkit to couple earth system models}, Environmental Modelling {\&}
  Software, 23, 1240--1249, \doi{10.1016/j.envsoft.2008.03.002}, 2008.

\bibitem[{Warner et~al.(2010)Warner, Armstrong, He, and Zambon}]{Warner2010}
Warner, J.~C., Armstrong, B., He, R., and Zambon, J.~B.: {Development of a
  Coupled Ocean–Atmosphere–Wave–Sediment Transport (COAWST) Modeling
  System}, Ocean Modelling, 35, 230--244, \doi{10.1016/j.ocemod.2010.07.010},
  2010.

\bibitem[{Wirtz and Kerimoglu(2016)}]{Wirtz2016}
Wirtz, K.~W. and Kerimoglu, O.: {Autotrophic Stoichiometry Emerging from
  Optimality and Variable Co-limitation}, Frontiers in Ecology and Evolution,
  4, \doi{10.3389/fevo.2016.00131}, 2016.

\bibitem[{Yakushev et~al.(2017)Yakushev, Protsenko, Bruggeman, Wallhead,
  Pakhomova, Yakubov, Bellerby, and Couture}]{Yakushev2017}
Yakushev, E.~V., Protsenko, E.~A., Bruggeman, J., Wallhead, P., Pakhomova,
  S.~V., Yakubov, S.~K., Bellerby, R. G.~J., and Couture, R.-M.: {Bottom RedOx
  Model (BROM v.1.1): a coupled benthic–pelagic model for simulation of water
  and sediment biogeochemistry}, Geoscientific Model Development, 10, 453--482,
  \doi{10.5194/gmd-10-453-2017}, 2017.

\bibitem[{Z{\"{a}}ngl et~al.(2015)Z{\"{a}}ngl, Reinert, R{\'{i}}podas, and
  Baldauf}]{Zaengli2015}
Z{\"{a}}ngl, G., Reinert, D., R{\'{i}}podas, P., and Baldauf, M.: {The ICON
  (ICOsahedral Non-hydrostatic) modelling framework of DWD and MPI-M:
  Description of the non-hydrostatic dynamical core}, Quarterly Journal of the
  Royal Meteorological Society, 141, 563--579, \doi{10.1002/qj.2378}, 2015.

\end{thebibliography}

\clearpage

\clearpage

\begin{table}
\caption{Components currently integrated in MOSSCO and described shortly in this paper.  Several other components 
are under development and not listed here.}
\label{tab:components}
\begin{tabular}{llr}
\hline
Pelagic ecosystem & \texttt{fabm\_pelagic\_component} & \refsec{fabm_pelagic}\\
Soil ecosystem & \texttt{fabm\_sediment\_component} & \refsec{fabm_sediment}\\
1D hydrodynamics & \texttt{gotm\_component} & \refsec{gotm}\\
3D hydrodynamics & \texttt{getm\_component} & \refsec{getm}\\
Filtration & \texttt{filtration\_component} & \refsec{filtration}\\
Erosion/sedimentation & \texttt{erosed\_component} & \refsec{erosed}\\
Wind waves & \texttt{simplewave\_component} & \refsec{waves}\\
Generic output & \texttt{netcdf\_component} & \refsec{output}\\
Generic input & \texttt{netcdf\_input\_component} & \refsec{input}\\
Link connector & \texttt{link\_connector} & \refsec{linkcopy}\\
Copy connector & \texttt{copy\_connector} & \refsec{linkcopy}\\
Nudge connector & \texttt{nudge\_connector} & \refsec{linkcopy}\\
Tracer transport & \texttt{transport\_connector} & \refsec{transport}\\
Benthic--pelagic coupling & \texttt{soil\_pelagic\_connector} & \refsec{bpcoupler}\\
 & \texttt{pelagic\_soil\_connector} & \refsec{bpcoupler}\\
\hline
\end{tabular}
\end{table}

\begin{table}
\caption{Acronyms and model abbreviations used in the text.}
\label{tab:abbreviations}
\scriptsize
\include{table_02_abbreviations}
\end{table}

\end{document}